# Parity-time symmetry in optical microcavity systems

Jianming Wen[1*], Xiaoshun Jiang[2,3,*], Liang Jiang[4*], and Min Xiao[2,3,5*]

[1]Department of Physics, Kennesaw State University, Marietta, Georgia 30060, USA
[2]National Laboratory of Solid State Microstructures, College of Engineering and Applied Sciences, and School of Physics, Nanjing University, Nanjing 210093, China
[3]Synergetic Innovation Center in Quantum Information and Quantum Physics, University of Science and Technology of China, Hefei, Anhui 230026, China
[4]Department of Applied Physics and Yale Quantum Institute, Yale University, New Haven, Connecticut 06511, USA
[5]Department of Physics, University of Arkansas, Fayetteville, Arkansas 72701, USA
*Email: jwen2@kennesaw.edu; jxs@nju.edu.cn; liang.jiang@yale.edu; mxiao@uark.edu.

Canonical quantum mechanics postulates Hermitian Hamiltonians to ensure real eigenvalues. Counterintuitively, a non-Hermitian Hamiltonian, satisfying combined parity-time (PT) symmetry, could display entirely real spectra above some phase-transition threshold. Such a counterintuitive discovery has aroused extensive theoretical interest in extending canonical quantum theory by including non-Hermitian but PT-symmetric operators in the last two decades. Despite much fundamental theoretical success in the development of PT-symmetric quantum mechanics, an experimental observation of pseudo-Hermiticity remains elusive as these systems with a complex potential seem absent in Nature. But nevertheless, the notion of PT symmetry has highly survived in many other branches of physics including optics, photonics, AMO physics, acoustics, electronic circuits, material science over the past ten years, and others, where a judicious balance of gain and loss constitutes a PT-symmetric system. Here, although we concentrate upon reviewing recent progress on PT symmetry in optical microcavity systems, we also wish to present some new results that may help to accelerate the research in the area. Such compound photonic structures with gain and loss provide a powerful platform for testing various theoretical proposals on PT symmetry, and initiate new possibilities for shaping optical beams and pulses beyond conservative structures. Throughout this article there is an effort to clearly present the physical aspects of PT-symmetry in optical microcavity systems, but mathematical formulations are reduced to the indispensable ones. Readers who prefer strict mathematical treatments should resort to the extensive list of references. Despite the rapid progress on the subject, new ideas and applications of PT symmetry using optical microcavities are still expected in the future.

1. Introduction
2. PT symmetry in quantum mechanics: Basic concepts and properties
    2.1   Parity and time reversal operators
    2.2   PT symmetry versus pseudo-Hermiticity
    2.3   Exceptional points and phase transition
3. PT symmetry in optical microcavity systems
    3.1   Paraxial optics versus quantum mechanics
    3.2   PT symmetry in coupled optical microcavities
    3.3   Proposal on anti-PT symmetry in optical microcavities
    3.4   On-chip optical nonreciprocity
4. Applications of PT-symmetric optical microcavities
    4.1   Single-mode microring lasers
    4.2   Coherent perfect laser-absorber
    4.3   Supersensitive sensing
5. Conclusion and prospects

1. Introduction

Optical microcavities (or microresonators) [1-3], confining light to small volumes by resonant recirculation, has become a distinct research area in photonics over the past two decades. Both physical implementations of these devices and their applications are highly differentiated, ranging from optical communications, cavity quantum electrodynamics (QED), nonlinear optics, novel laser sources, bio-sensing, quantum optics, quantum information processing, to integrated photonics. Although an ideal cavity would trap light indefinitely at resonant frequencies with exact values, in practice experiments with photonics are intrinsically non-Hermitian owing to gain and loss. As a result, the resonance nature of the interaction of light with cavities does not always bring about desired enhancement on optical properties of interest only. Recent experimental progress on microcavities with alternating gain and loss in a balanced manner [4-9], however, shines a new light to the problem. In these demonstrations [4-9], light wave is typically amplified in some regions of the system and is attenuated in others. Heuristically, physics seems only limited to the compensation of loss by equal amount of amplification. On the contrary, it turns out that subject to this condition, a nontrivial, sharp phase transition arises further from the evolution of characteristics of supermode profiles with only real frequency components to with complex spectra. Alternatively, on crossing a unique "exceptional point" (EP) [10-16] in parameter space, the eigen-spectra cease to be real and the underlying modes coalesce to a single mode, because of the nontrivial wave interference. In fact, such structures share a common feature, that is, they belong to open but parity-time (PT) symmetric systems [17-28]. The PT symmetry in optics [20-27] amounts to the necessary condition $n(\boldsymbol{r}) = n^*(-\boldsymbol{r})$ with $n(\boldsymbol{r})$ the complex refractive index of the medium, which demands the real part of the refractive index be an even function of position whereas the imaginary part be odd.

Historically, interest in PT symmetry originated from the effort on the generalization of standard quantum mechanics by the inclusion of non-Hermitian Hamiltonians [17-19,29-32]. Although such extension of traditional quantum theory is highly controversial, Bender and Boettcher [33] theoretically showed that a class of non-Hermitian Hamiltonians, commuting with the joint parity and time-reversal operator, can possess entirely real spectra above a phase-transition point (or EP). Below that critical point, PT symmetry spontaneously breaks down and the Hamiltonians start to undergo a phase transition that admits complex eigenvalues instead. This counterintuitive discovery, on the one hand, radically challenges one of the basic axioms in canonical quantum mechanics formulated by Dirac and von Neumann that the Hermiticity of an operator guarantees real eigenvalues and the orthonormality between eigenstates with different eigenvalues for closed physical systems. On the other hand, it dramatically shapes our cognition on open systems, where it is commonly believed to have complex eigenvalues and non-orthogonal eigenfunctions. The significance of this pioneering work by Bender and Boettcher has immediately aroused considerable theoretical effort [17-19,29-32] in extending Hermitian quantum theory to non-Hermitian but PT-symmetric operators. In the past two decades, an intensive research has been dedicated to the exploration of a more general class of pseudo-Hermitian operators [11,20,25] with special symmetries and purely real eigenvalues. The developments on this front have helped to establish an active research field, PT-symmetric quantum mechanics, which has been covered in a series of reviews [19,29,34] and special issues [30-22]. Despite the impressive theoretical success in prospering PT-symmetric quantum mechanics, a viable experimental observation of such pseudo-Hermiticity remains yet elusive in real physical settings.

In addition to quantum mechanics, the scope of PT symmetry has been rapidly expanded to a variety of physics branches. The first attempt on constructing an optical analog was made by Ruschhaupt *et al* [35], who theoretically analyzed light propagation in a medium with an even refractive index and odd gain-loss landscape. However, it was the works by Christodoulides and his coworkers [36-38] that exceedingly stimulated the attention on PT symmetry in the realm of optics and photonics. By noticing the mathematical isomorphism between the quantum Schrödinger equation and the paraxial optical diffraction equation, they [36] suggested to realize complex PT-symmetric potentials through judiciously making use of refractive indices with balanced gain and loss in optical settings. In particular, with coupled waveguides they formulated PT-symmetric optics by providing a simple but nontrivial framework for the study of PT-symmetric systems. It was shown that in waveguide structures, the parity operator leads to spatial reflection, while the time-reversal operator reverses the propagation direction. One of the most striking PT properties stems from the appearance of a sharp, symmetry-breaking transition, once a non-Hermitian parameter crosses a certain threshold. This transition signifies the occurrence of a spontaneous PT symmetry breaking from the exact- to the broken PT phase in a classical way. Such a peculiar behavior is not only of interest to fundamental research but also gives rise to a breadth of new opportunities for flexible light control and manipulation. Subsequent experiments [39,40] have indeed confirmed those theoretical predictions. Inspired by these optical realizations, the extension of PT symmetry to other branches of physics [4-9,42-66] is then readily followed and has revealed numerous intriguing phenomena [67-130] that may be inaccessible with usual Hermitian arrangements. Moreover, by incorporating nonlinearity [22-26] one can largely enrich the overall dynamics with numerous exotic phenomena beyond the capability of conservative architectures.

Indeed, work on PT symmetry is very diverse and has resulted in a long list of publications that run the gamut from reports of its experimental realizations, to discussions of its fundamental physics, to suggestions for implementation variations dictated by practical considerations. In this review, we aim at presenting a compact overview, but by no means a comprehensive discussion, of recent advances on PT-symmetric optical microcavities with emphasis on some intriguing optical phenomena and new emerging applications. We notice that PT symmetry has already been summarized in a series of reviews [18-28] from different perspectives. Different from these existing reviews, however, throughout this article there is an effort to clearly present the physical aspects of the PT-symmetry. Mathematical formulations are reduced to the indispensable ones. Readers who prefer rigorous mathematical treatments should resort to the extensive list of references. We in the meanwhile wish to present some new results and observations that have not yet been discussed elsewhere. It is hoped that these new additions may further accelerate and expand the research interest and scopes in PT-symmetric optical microcavities.

Since PT symmetry originated from extending standard quantum mechanics into complex plane, in Section 2 we being our review with a short survey on its historical development by summarizing only the major characteristics associated with. To make the argument more meaningful, a two-level toy model will be introduced in the end to provide an intuitive understanding of the abstract concepts. As it will become clear throughout the whole content, it is this toy model that essentially paves the way for classical emulations with a variety of physical settings.

Section 3 is devoted to recent observations of PT symmetry in optics, especially optical microcavities. We first follow the historical development by establishing the quantum-classical analogy between the Schrödinger equation for a quantum particle and the paraxial propagation equation of light in a medium, which enables the first experimental implementations in coupled waveguides in the spatial domain. We then focus on recent advancement of achieving PT/anti-PT symmetry with use of optical microcavities in the temporal domain. Owing to their flexible optical properties through design, optical microcavities also allow us to build connections with other research areas, which include optical nonreciprocity and cavity optomechanics. In this section, we further present a theoretical proposal for the realization of anti-PT symmetry using three microcavities. Of interest, this structure fulfills the accomplishment of both PT and anti-PT symmetry within in the same architecture.

In Section 4 the potential applications of PT-symmetric optical microcavities are concentrated on few recent experimental demonstrations. In comparison with traditional structures, PT symmetry brings novel functionalities, and opens new prospects for the development of novel photonic devices such as single-mode lasers and supersensitive sensors.

Finally, we conclude our review with a brief summary and outlook in Section 5.

2. PT symmetry in quantum mechanics: Basic concepts and properties

We begin, in this section, by briefly overviewing the essential concepts in the theory of PT-symmetric quantum mechanics. We do not intend to record all available materials of this extremely broad field, but rather to focus on the critical ideas that are relevant to the content in the subsequent sections. For comprehensive reviews on non-Hermitian operators in physics and mathematics, besides the works [10-14,17-19,29-34] listed above, interested readers please refer to the reviews [131-135], as well as the monograph of Moiseyev [136].

### 2.1 Parity and time-reversal operators

Standard quantum mechanics demands Hermitian Hamiltonians, to describe closed physical systems, to ensure real eigenspectra and the orthonormality between eigenstates with different eigenvalues. As one axiom in the mathematical framework of conventional quantum theory, the postulation on Hermiticity guarantees real observables and probability-preserving time evolution (or unitarity). In contrast, the dynamics of open systems is typically described by non-Hermitian Hamiltonians. Owing to the seemingly impossible conservation of energy, complex eigenvalues are commonly taken for granted in non-conservative systems. Since the early days of quantum mechanics, considerable efforts have been dedicated to incorporating non-Hermitian Hamiltonians into the well-accepted Hermitian representations. Throughout the enterprise, early notable achievements include pioneering works by Gamow [137], Feshbach, Porter and Weisskopf [138], and Lindblad [139,140]. Indeed, these successful developments lead to a more formal basis for describing the dynamics of certain open quantum systems [141-143]. In spite of these achievements, yet the interpretation behind is not straightforward all the time. As such, Barton in his book *Introduction to Advanced Field Theory* [144] used to comment: "A non-Hermitian Hamiltonian is unacceptable partly because it may lead to complex energy eigenvalues, but chiefly because it implies a non-unitary $S$ matrix, which fail to conserve probability and makes a hash of the physical interpretation."

Although the research on non-Hermitian generalization of quantum mechanics has a long history, for a non-Hermitian Hamiltonian to have real eigenspectra was due to the theoretical work by Wu in 1959 [145], who first used an anharmonic oscillator (with mass $m$) associated with a pure imaginary cubic potential

$$\widehat{H} = \frac{\hat{p}^2}{2m} + i\hat{x}^3, \tag{1}$$

with the momentum operator $\hat{p} = -i\hbar \frac{d}{dx}$ and the position operator $\hat{x}$, to analyze the ground state of a quantum system of hard spheres. Quite surprisingly, he found that the eigenvalues were real and no longer divergent. This counterintuitive finding made a quite stir in the community and stimulated many follow-up studies. For example, it was later further examined in the Reggeon field theory [146,147], the Yang-Lee edge singularity [148-150], and the perturbation theory of odd anharmonic oscillators [151]. Another important contribution in the development of non-Hermitian quantum mechanics was the publication by Haydock and Kelly in 1975 [152], where they pointed out for the first time that while Hermiticity was sufficient to ensure real eigenvalues, it was not necessary and sufficient. The significance of their work is that they put forward a very interesting question about the identification of a necessary and sufficient condition for real eigenvalues. The next big step forward was a justification for the use of complex eigenvalues by Hatano and Nelson in 1997 [153], where they applied depinning field lines of a non-Hermitian external magnetic field to a type II semiconductor. Nevertheless, in a long period no one could find a physical reason why such a non-Hermitian Hamiltonian allows all real eigenvalues. At the same time, undoubtedly, objections also existed in the community, as these people more concerned about the consequences of the results. Did the probabilistic interpretation of quantum mechanics still hold? Was the norm positive and conserved? All kinds of fundamental questions were raised in regard to introducing non-Hermitian Hamiltonians into Hermitian quantum theory.

The truth had to wait until 1998. In that year, Bender and Boettcher [33] made a seminal discovery by showing that Hermiticity is a sufficient but not necessary condition for spectral reality and unitary time evolution. In particular, they found a broad class of non-Hermitian Hamiltonians which can have a set of eigenstates with real spectra, provided they commute with the combined parity operator $\hat{P}$ and time-reversal operator $\hat{T}$,

$$\hat{P}\hat{T}\hat{H} = \hat{H}\hat{P}\hat{T} \Leftrightarrow [\hat{P}\hat{T}, \hat{H}] = 0, \tag{2}$$

a necessary condition. As two fundamental discrete symmetries in physics, $\hat{P}$ and $\hat{T}$ are, respectively, defined by their actions on the dynamical variables $\hat{x}$ and $\hat{p}$. Specifically, the linear parity operator $\hat{P}$ has the effect of flipping the sign of $\hat{x}$ and $\hat{p}$: $\hat{x} \to -\hat{x}$ and $\hat{p} \to -\hat{p}$; while the antilinear time-reversal operator $\hat{T}$ [154] applies the operations of $\hat{x} \to \hat{x}$, $\hat{p} \to -\hat{p}$, and $i \to -i$. Here, $\hat{T}$ changes the sign of $i$ due to the requirement on preserving the fundamental commutation relation $[\hat{x}, \hat{p}] = i\hbar$ in traditional quantum mechanics. In addition, $\hat{P}$ and $\hat{T}$ are unitary and have the following properties,

$$\hat{P}^2 = \hat{T}^2 = \hat{I} \text{ and } [\hat{P}, \hat{T}] = 0, \tag{3}$$

where $\hat{I}$ is the identity operator. Now multiplying by $\hat{T}\hat{P}$ from the right-hand side in Eq. (2) gives

$$\hat{P}\hat{T}\hat{H}\hat{T}\hat{P} = \hat{H}, \tag{4}$$

with the help of Eq. (3). Equation (4) is PT invariant as $(\hat{T}\hat{P})^\dagger = \hat{P}\hat{T}$ (the symbol † here stands for the Hermitian conjugate). In the Schrödinger equation, for Hamiltonians of the form

$$i\hbar \frac{\partial}{\partial t}\Psi = \hat{H}\Psi, \ \hat{H} = \frac{\hat{p}^2}{2m} + V(x) \tag{5}$$

with $V(x)$ the (complex) potential energy of the system, the necessary condition (2) immediately implies that the real part of $V(x)$ be an even function of the coordinate and the imaginary part be an odd function, i.e.,

$$V(x) = V^*(-x) \tag{6}$$

with ∗ denoting the complex conjugate.

2.2 PT symmetry versus pseudo-Hermiticity

Unlike Hermiticity, PT symmetry is not a sufficient condition to guarantee the spectral reality. When combined with the requirement of the unbroken PT symmetry, however, it becomes sufficient. To illustrate this, let $E_j$ be an eigenvalue of $\hat{H}$ with respect to the eigenvector $|\Psi_j\rangle$ (which also simultaneously becomes the eigenvector of $\hat{P}\hat{T}$ with an eigenvalue λ). That is,

$$\hat{H}|\Psi_j\rangle = E_j|\Psi_j\rangle, \tag{7}$$

Applying the PT transformation (4) to Eq. (7) leads to

$$\hat{P}\hat{T}\hat{H}\hat{T}\hat{P}|\Psi_j\rangle = \hat{P}\hat{T}E_j\hat{T}\hat{P}|\Psi_j\rangle.$$

As $\hat{T}E_j\hat{T} = E_j^*$ and $\hat{P}^2 = \hat{I}$ (3), the above equation yields

$$\hat{P}\hat{T}\hat{H}\hat{T}\hat{P}|\Psi_j\rangle = \hat{H}|\Psi_j\rangle = E_j|\Psi_j\rangle = E_j^*|\Psi_j\rangle,$$

where Eqs. (4) and (7) have been used. Hence, $E_j = E_j^*$ and the eigenvalue $E_j$ is real. Because the procedure is applicable to every eigenvalue of $\hat{H}$, this concludes the realness of its spectra. The above proof fully relies on the crucial assumption on $|\Psi_j\rangle$ to be an eigenvector of both $\hat{H}$ and $\hat{P}\hat{T}$. In quantum mechanics, we learn that if a symmetry transformation is represented by a linear operator $\hat{A}$ and if $\hat{A}$ commutes with $\hat{H}$, the eigenstates of $\hat{H}$ are then also eigenstates of $\hat{A}$. Unfortunately, because of the antilinearity of the joint operator $\hat{P}\hat{T}$, one has to make the extra but nontrivial assumption that the PT symmetry of $\hat{H}$ is unbroken. What happens if this assumption (spontaneously) breaks down? It turns out that there will be a striking phase transition associated with the presence of complex eigenvalues in the spectra of $\hat{H}$. Different from Hermitian quantum mechanics, PT symmetry does not promise the completeness and orthomornality of eigenvectors of the operator. It is worth to emphasize that acceptable complex Hamiltonians may be either Hermitian $\hat{H} = \hat{H}^\dagger$ in the Dirac sense or PT-symmetric, but not both; in contrast, real symmetric Hamiltonians can be both Hermitian and PT-symmetric. From this point of view, the central idea of PT-symmetric quantum mechanics [17-19,33,155-158] is to relax the strong condition of Hermiticity for a quantum Hamiltonian with the weaker condition that it have space-time reflection symmetry. Alternatively, the proposal on PT-symmetric Hamiltonians does not give up Hermiticity. Rather, it offers the possibility of studying new quantum theories that might even describe measurable physical phenomena in the physical world.

Using PT symmetry as an alternative condition to Hermiticity, one can devise infinitely new Hamiltonians that would have been rejected since the early days of quantum mechanics, simply because they are apparently non-Hermitian. The Hamiltonian (1) turns out to be one such example. Based upon this Hamiltonian, a more general class [33] of PT-symmetric ones can be readily deduced by utilizing the fact that any real function of $i\hat{x}$ is PT-symmetric:

$$\hat{H} = \frac{\hat{p}^2}{2m} + \hat{x}^2(i\hat{x})^\varepsilon, \ \varepsilon \in \text{Real}. \tag{8}$$

When $\varepsilon = 0$, the Hamiltonian (8) reduces to that of the classic harmonic oscillator whose exact solutions are addressed in most of quantum mechanics textbooks. As illustrated in their numerical studies, Bender and Boettcher [33] observed that the spectra of $\hat{H}$ (8) exhibit three distinct behaviors with an abrupt phase transition as a function of $\varepsilon$ (see figure 1). Specifically, for all $\varepsilon \geq 0$ the corresponding Hamiltonians (8) are in the regime of unbroken PT symmetry, and their eigenspectra are infinite, discrete, entirely real and positive. Obviously, the Hamiltonian (1) is just one member of this huge class of complex Hamiltonians. This explains why the Hamiltonian (1) possesses all real, positive, and discretized energy levels. For $-1 < \varepsilon < 0$, there are only a finite number of positive real eigenvalues but an infinite number of complex conjugate pairs of eigenvalues. In this region, the PT symmetry spontaneously breaks down. As $\varepsilon$ decreases from 0 to −1, adjacent energy levels merge into complex conjugate pairs at the high end of the spectra. Eventually, the only remaining real eigenvalue is the ground-state energy but diverges till $\varepsilon \to -1^+$. When $\varepsilon \leq -1$ no real spectrum is available. As we can see from Fig. 1, at $\varepsilon = 0$ lies the conventional Hermitian Hamiltonian of the quantum-mechanical harmonic oscillator, which marks the phase-transition point between the unbroken and broken PT regions. Rooted in the above features, the term "PT symmetry" is coined to describe these new non-Hermitian complex Hamiltonians but having real energy levels.

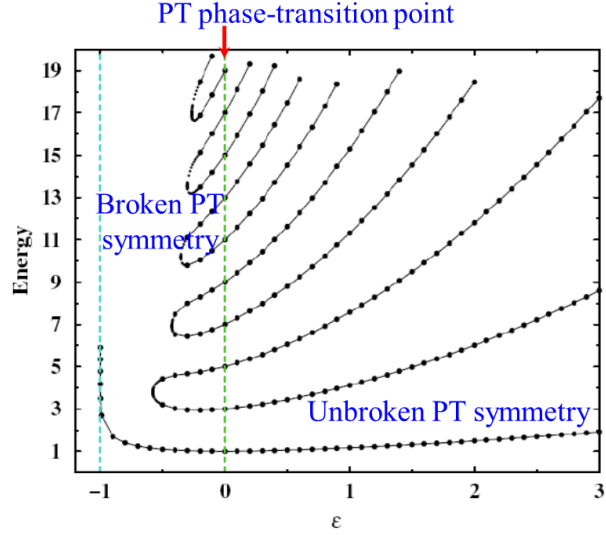

Figure 1 Real, positive energy levels of the Hamiltonians (8) as a function of the real parameter $\varepsilon$. For $\varepsilon \geq 2$, the entire spectra are definitely real and positive. For $-1 < \varepsilon < 0$, the spectra have a finite number of real positive eigenvalues but have unlimited complex conjugate pairs of eigenvalues. In this regime, the number of real eigenvalues decreases until only the remaining ground-state energy, along with the reduction of $\varepsilon$. As $\varepsilon \to -1^+$, the ground-state energy diverges. For $\varepsilon \leq -1$ no real eigenenergy can be attainable. (Reproduced from Ref. [33])

For a non-Hermitian Hamiltonian to possess fully real eigenspectra, a sufficient and necessary condition, in fact, can be mathematically formulated in terms of a more general property called pseudo-Hermiticity [34,159]. In the framework of pseudo-Hermitian quantum mechanics [29,159-165], PT symmetry does not play a basic role and no further need introduces a C-operator [156,166] to make the theory well-defined, although they can be nevertheless defined in general. On the contrary, all that is needed is to determine the class of non-Hermitian Hamiltonians that are capable of generating a unitary evolution and a procedure that associates, to each member of this class, a positive-definite inner product that renders it Hermitian (or self-adjoint). There are always an infinite class of positive-definite inner products meeting this condition. Any inner product may be defined by a certain linear metric operator, $\hat{\eta}$, which determines the kinematics of pseudo-Hermitian quantum systems. The Hamiltonian operator $\hat{H}$ is linked to $\hat{\eta}$ via the $\hat{\eta}$-pseudo-hermiticity relation (sometimes also called the similarity transformation),

$$\hat{H}^\dagger = \hat{\eta}\hat{H}\hat{\eta}^{-1}, \text{ and } \hat{\eta} = \hat{U}^\dagger\hat{\eta}\hat{U}, \tag{9}$$

where $\hat{U}$ is a unitary operator. Clearly, when $\hat{\eta} = \hat{I}$, the pseudo-Hermitian Hamiltonian coincides with a Hermitian one. This simple observation indicates that Hermitian Hamiltonians define a subset of pseudo-Hermitian Hamiltonians. Moreover, it is not difficult from Eq. (9) to see that pseudo-hermiticity is unitary invariant.

By defining the automorphism as an invertible operator that maps an inner-product space onto itself, Mostafazadeh [159] proved that if $\hat{H}$ is a non-Hermitian Hamiltonian with a discrete spectrum and has a complete biorthonormal eigenbasis, then $\hat{H}$ is pseudo-Hermitian if and only if one of the following conditions hold: (i) the spectrum of $\hat{H}$ is real, or (ii) the complex eigenvalues come in complex conjugate pairs and the multiplicity of the eigenvalue pairs is the same. The characteristic transform of the spectra from real to complex entails a nontrivial phase transition. The PT phase transition is thus naturally incorporated into the framework of pseudo-Hermitian quantum mechanics. To ensure a real spectrum, pseudo-hermiticity is required but is not a sufficient condition on its own. To formulate a necessary and sufficient condition, Mostafazadeh [160] then introduced another theorem: "Let $\hat{H}$ be a Hamiltonian that acts in a Hilbert space, has a discreate spectrum, and allows a complete set of biorthonormal eigenvectors $\{|\Psi_j\rangle, |\Phi_j\rangle\}$. Then the spectrum of $\hat{H}$ is real if and only if there is an invertible linear operator $\hat{O}$ so that $\hat{H}$ is $\hat{O}\hat{O}^\dagger$-pseudo-Hermitian."

One interesting question is whether (C)PT-symmetry is in fact a subclass of pseudo-hermiticity. Mostafazadeh [161] claims that it is. He went on to prove that all diagonalizable linear operators with a discrete spectrum will have an antilinear symmetry, leading him to the following two corollaries: (i) any diagonalizable linear operator that possesses a discrete spectrum (real eigenvalues) is anti-pseudo-Hermitian; and (ii) any diagonalizable pseudo-Hermitian linear operator that possesses a discrete spectrum will have antilinear symmetry. Since PT symmetry is an antilinear symmetry, the PT-symmetric Hamiltonians imply that $\hat{\eta}$-pseudo-hermiticity is present, $\hat{\eta} = \hat{\tau}\hat{P}\hat{T}$ with $\hat{\tau}$ an antilinear operator obeying the relation $\langle\Phi|\hat{\tau}|\Psi\rangle = \langle\Psi|\hat{\tau}|\Phi\rangle$. In addition, Mostafazadeh [164] demonstrated the unitary equivalence between PT-symmetric quantum mechanics and Hermitian quantum mechanics, and the recovery of Hermitian Hamiltonians from every PT-symmetric Hamiltonian.

Though pseudo-Hermitian quantum mechanics shows abundant (mathematical) properties beyond PT-symmetric one, from now on we will not discuss it as the focus of this article is PT symmetry. Two comprehensive reviews on pseudo-Hermitian quantum mechanics have recently been given by Mostafazadeh [29,165], where an extensive list of references can be found. Another attention that should be paid is that the content of PT symmetry discussed in this subsection is more about a quantum particle subject to canonical continuous variables $(\hat{x}, \hat{p})$. The behavior associated with such a Hamiltonian can be qualitatively different from that with a Hamiltonian confined in a finite-dimensional Hilbert space, as we shall see in the following subsection.

### 2.3 Exceptional points and phase transition

Non-Hermitian quantum mechanics has interesting applications to open quantum systems. In particular, non-Hermitian theories can be used as tools to model the behavior of open quantum systems. For simplicity, from now on we will use a symbol-without-hat to denote an operator for a matrix representation, unless otherwise specified. One of the benefits of doing this is that the notions are appropriate for both quantum and classical treatments without adding confusion. To make the arguments more illustrative, as a two-state toy model, let us look at a simple $2 \times 2$ matrix Hamiltonian [156]

$$H = \begin{pmatrix} re^{i\theta} & s \\ s & re^{-i\theta} \end{pmatrix} \neq H^\dagger \quad (\{r,s\} \in \text{Real and } \theta \in [0,2\pi)), \tag{10}$$

which is generally not Hermitian. But it is PT symmetric, where the parity operator is (equivalent to the Pauli matrix $\sigma_1$)

$$P = \begin{pmatrix} 0 & 1 \\ 1 & 0 \end{pmatrix} = P^{-1} = P^\dagger, \tag{11}$$

and the time-reversal operator $T$ corresponds to complex conjugation. Note that the parity operator (11) satisfies $P = P^{-1} = P^\dagger$.

After a little algebra, one can easily obtain its two eigenvalues

$$\epsilon_\pm = r\cos\theta \pm \sqrt{s^2 - r^2\sin^2\theta}. \tag{12}$$

Evidently from Eq. (12), there are two parametric regions to consider, one for which the square root is real and the other for which it is imaginary. That is, when $s^2 < r^2\sin^2\theta$, the energy eigenvalues form a complex conjugate pair, which indicates the region of broken PT symmetry. While for $s^2 > r^2\sin^2\theta$, the eigenvalues are real, implying the region of unbroken PT symmetry. In this unbroken region, the simultaneous eigenstates of both $H$ and $PT$ have the form

$$|\psi_{\epsilon_+}\rangle = \frac{1}{\sqrt{2\cos\alpha}}\begin{pmatrix} e^{i\alpha/2} \\ e^{-i\alpha/2} \end{pmatrix} \text{ and } |\psi_{\epsilon_-}\rangle = \frac{i}{\sqrt{2\cos\alpha}}\begin{pmatrix} e^{-i\alpha/2} \\ -e^{i\alpha/2} \end{pmatrix}, \tag{13}$$

with $\alpha = \sin^{-1}\left(\frac{r\sin\theta}{s}\right)$. Because the eigenvectors are generally biorthogonal for an open system [167], the eigenstates (13) here obey the orthonormalization condition $\langle\psi_{\epsilon_\pm}|\psi_{\epsilon_\pm}\rangle = \delta_{+-}$. If $s^2 < r^2\sin^2\theta$, the states (13) are no longer eigenstates of $PT$ because $\alpha$ becomes imaginary. Moreover, the PT norm of the energy eigenstate vanishes. $s^2 = r^2\sin^2\theta$ is the exact point where the phase transition occurs between broken and unbroken symmetry. At this branch point, the Hamiltonian (10) turns into a nondiagonal Jordan block. As a result, the two eigenvalues collide and the two eigenvectors are linearly dependent. In other words, the respective algebraic multiplicity of the eigenvalue is two, larger that the geometric multiplicity of one. In fact, this PT-symmetry breaking point reflects all the characteristics of an exceptional point singularity. In general, exceptional points appear as singularities of non-Hermitian eigenvalue problems and can be compared with degeneracies in Hermitian operators. The exceptional point is intrinsically different from Hermitian degeneracies, conforming to the situation in which both the eigenvalues and eigenvectors coalesce. In the limit $\theta \to 0$ in this two-state system, the Hamiltonian (10) becomes Hermitian.

Interestingly, for the toy model presented above, one can introduce the counterpart of PT symmetry, "anti-PT symmetry", which instead anti-commutes with the Hamiltonian, $\{PT, H_{\text{anti}}\} = 0$. Mathematically, the Hamiltonian (10) would become anti-PT symmetric on multiplying by $i$ [64],

$$H_{\text{anti}} = \begin{pmatrix} ire^{i\theta} & is \\ is & ire^{-i\theta} \end{pmatrix}. \tag{14}$$

Based on this observation, the concept of anti-PT symmetry was explicitly proposed and demonstrated in the work of Peng *et al* [64]. The mathematical transformation between $H$ and $H_{\text{anti}}$, as a consequence, implies immediately that the behavior of an anti-PT system would be conjugate to that of a PT system. In addition, the action of $PT$ on the Hamiltonian (10) returns $H$ itself while for the Hamiltonian (14), the same operation yields $-H_{\text{anti}}$ as a contrast. One can quickly check that indeed, $H_{\text{anti}}$ anti-commutes with the joint $PT$ operator defined in (11). Without difficulty, the two eigenvalues can be readily obtained with the form

$$\mathcal{E}_\pm = ir\cos\theta \pm i\sqrt{s^2 - r^2\sin^2\theta}. \tag{15}$$

But the eigenstates remain unchanged and coincide with the states (13). By comparing Eq. (15) with Eq. (12), one may notice that $\mathcal{E}_\pm = iE_\pm$, as expected from the relationship $H_{\text{anti}} = iH$. Opposite to the PT case, for $s^2 < r^2\sin^2\theta$, the two eigenvalues are complex and the system is termed in the regime of unbroken anti-PT symmetry; for $s^2 > r^2\sin^2\theta$, the pair of eigenvalues become purely imaginary, corresponding to the region of anti-PT symmetry breaking. The anti-PT phase transition point stays at $s^2 = r^2\sin^2\theta$. Distinct from the PT case, the anti-PT Hamiltonian (14) always maintain non-Hermitian irrespective of the value of $\theta$. At $\theta = \left\{\frac{\pi}{2}, \frac{3\pi}{2}\right\}$ emerges perfect anti-PT symmetry in the sense that $\mathcal{E}_\pm$ are either real or purely imaginary.

This sort of toy models have been extensively studied in the literature, owing to their simplicity. The involved research has generated hundreds of theoretical publications in a variety of journals. These include the work by Bender and his colleagues [156] where they conceived a charge operator $C$ so that the inner product of PT symmetry can be properly defined through the co-application of $CPT$. Even though the theory looks attractive, in this article we will not go into further discussions here. Readers interested in this type of research may take advantage of the cited references for additional reading.

Before moving to the next section, we wish to make the following few remarks on PT-symmetric quantum mechanics. First of all, although the toy model undoubtedly serves as an intuitive way to illustrate the essence of PT and anti-PT symmetry, due to the lack of physical substance, so far its practical implementation remains highly elusive in terms of the demanding complex PT-potentials. Second, unlike the case with dynamical variables (see Section 2.2), the toy model here pertains to a discrete version. As shall be evident in the following sections, it is this toy model that has significance in optics and other branches of physics where complex PT potentials can be easily realized by properly balancing gain and loss. Third, as will become clear in the rest content, each matrix element in the Hamiltonians (10) and (15) has the well-defined physical meaning for different topics. Consequently, this leads to a tight connection with other research areas.

3. PT symmetry in optical microcavity systems

Regardless of the impressive theoretical success in the development of non-Hermitian quantum domain, yet, the lack of any experimental support makes these theories highly skeptical as physical theories in reality. Notwithstanding that, in the past decade non-Hermitian systems with the notion of PT symmetry in particular, as an emerging platform of interdisciplinary research among optics [39,40], photonics [47-49,53,54,59,65,66], AMO physics [63,64], acoustics [60-62], microwave mechanics [44,45], electronic circuits [41-43], and material science [46,50-52,55-58], have attracted extensive studies and initiated numerous intriguing prospects [67-130] beyond the capabilities of conservative structures. Rapid progress in its rich interplay with other physical phenomena has exceedingly reshaped the original intention to be the paradigm for extending Hermitian quantum mechanics. This section is committed to recent advances in experimenting PT symmetry using optical microcavities. In order to make the story self-contained, we start with the course on establishing PT-symmetric optics through the formal equivalence between the paraxial optical propagation equation in a medium and the Schrödinger equation for a quantum particle. We then continue to review recent experimental demonstrations of PT symmetry utilizing optical microcavities. As tunable photonic PT systems can add new exciting functionalities being potentially able to find even broader applications over the field of photonics, in the rest part of this section we will pay attention to few such efforts including on-chip optical nonreciprocity. Besides, we would like to present a new proposal on realizing both PT and anti-PT in the same microcavity setting.

3.1 Paraxial optics versus quantum mechanics

Quantum physics are fundamentally different from classical physics even from a conceptual point of view. A well-known example is that, in quantum theory the wavefunction is a complex probability amplitude, while in classical optics, its analog – the electromagnetic field – is a measurable real quantity. In fact, since the birth of quantum mechanics, its founders made many attempts to find at least a formal connection to classical physics. Among these efforts, a notable view came from Schrödinger [170], who believed classical dynamics of a point particle be the *geometrical optics* approximation of a linear-wave equation, in the same as ray optics be a limiting approximation of wave optics. In the subsequent years some rigorous mathematical analogies have been progressively identified between classical optics and quantum mechanics. Among them, one of the best known and widely exploited is based upon the similarity between the time-independent Schrödinger equation and the time-independent Helmholtz equation. Historically, this analogy has successfully led to many fruitful discoveries and designs, such as i) designing multilayered optical structures, as their quantum counterparts with 0D, 1D, or 2D geometry [171], with the same transmission characteristics; ii) the transverse modes of aspherical laser resonators resembling the eigenstates of the stationary Schrödinger equation with a potential well determined by the mirror profile [172]; iii) the ladder operator description of the Hermit-Gauss or the Laguerre-Gauss modes of a laser beam equivalent to that of the quantum harmonic oscillator [173]; iiii) the mimic of a coherent state through the refraction of displaced light beams by lenses [174]; and v) the optical simulation [175] of the Franck-Condon principle as well as the Ramsauer-Townsend effect in the mismatch of a mode passing through two fibers with different refractive index distributions; to name just a few. An excellent account on the topic please refer to the textbook [176] written by Dragoman and Dragoman. In these analogies, the Helmholtz equation by the paraxial approximation is typically reduced to the time-independent Schrödinger equation, where the propagation distance, wavelength, and refractive index correspond to the time, Planck's constant, and potential in quantum mechanics. The momentum in the quantum world tallies with the ray direction in the optical case. As a result, ray optics is recovered in the limit as the wavelength approaches zero.

In 2007 this analogy was rigorously employed by Christodoulides and his coworkers [36] to construct PT symmetry within the realm of optics. More specifically, let us consider a 1D dielectric planar inhomogeneous medium having a relative permittivity that only varies along the $x$ axis, $\varepsilon(x) = n^2(x)$. Here, the complex refractive index $n(x) = n_R(x) + in_I(x)$ with $n_R(x)$ representing the real index profile of the structure, and $n_I(x)$ denoting the gain or loss component. Then the optical beam propagation within the medium can be theoretically described, in the paraxial diffraction approximation, as a Schrödinger-like equation with 1D by assuming the electric field $E = \phi(x,z)e^{i(kz-\omega t)}$,

$$i\frac{\partial \phi}{\partial z} = H_p \phi, \quad H_p = -\frac{1}{2k}\frac{\partial^2}{\partial x^2} - k\varepsilon(x), \tag{16}$$

with $k = 2\pi/\lambda_0$ and $\lambda_0$ being the vacuum wavelength of light. It shall be noted that Hermitian Hamiltonians are in line with the cases where the optical energy is conserved and $\varepsilon(x)$ (or $n(x)$) is real (see. Fig. 2(a)). In a continuous medium, the parity operator $P$ performs the spatial reflection but the time-reversal operator $T$ reverses the propagation direction. Similar as in the quantum case (6), in order for $\varepsilon(x)$ to be PT-symmetric, the condition $\varepsilon(x) = \varepsilon^*(x)$ must be fulfilled. This condition arises from the necessity for the Hamiltonian (16) to commute with the parity-time operator $PT$ in such a way that $\phi$ is a common eigenstate of both $H$ and $PT$. Practically, it demands that the optical complex potential $k\varepsilon(x)$ consist of a symmetric real refractive index guiding $n_R(x) = n_R(-x)$ and an antisymmetric gain/loss profile $n_I(x) = -n_I(-x)$. Or, $n(x) = n^*(-x)$ in a compact way (see Fig. 2(a)). By expanding the scalar field amplitude $\phi(x,z)$ onto its eigenvalue and corresponding eigenmode as $\phi(x,z) = u(x)e^{i\beta z}$, Eq. (16) reduces to the scalar Helmholtz equation

$$\left[\frac{1}{2k}\frac{\partial^2}{\partial x^2} + k\varepsilon(x)\right]u = \beta u. \tag{17}$$

The mode propagation constant $\xi$ is thus given by $\xi = k + \beta$. Note that Eq. (17) agrees with the stationary Schrödinger equation for a quantum particle under the formal substitution $1/2k \to -\hbar^2/2m$.

In contrast to the quantum case discussed in Section 2, one fundamental aspect associate with optical PT components (as well as others) has to do with their coupled-mode interactions. Abiding by this rule, a simpler arrangement for Eq. (17) is to look at a PT coupler with one component being optically pumped to provide gain $\gamma$ while the neighbor arm experiencing equal amount of loss. Under these conditions, one can plug the ansatz $\phi(x,z) = [a_1(z)u_1(x) + a_2(z)u_2(x)]e^{i\beta z}$ into the paraxial propagation equation (16) with the normalization of the eigenmode $u_j(x)$ according to $\int dx\, u_m^*(-x)u_j(x) = \delta_{mn}$. After straightforward algebra, one can thence show that the evolution of the system is governed by a set of coupled-mode theory [175,176],

$$i\frac{dA}{dz} = HA, \quad A = \begin{pmatrix} a_1 \\ a_2 \end{pmatrix}, \quad H = \begin{pmatrix} i\gamma & -\kappa \\ -\kappa & -i\gamma \end{pmatrix}, \tag{18}$$

where $\kappa$ is the mode coupling coefficient. Recall the toy model sketched in Section 2.3. It is not difficult to find that the non-Hermitian Hamiltonian in Eq. (18) is genuinely PT symmetric, if setting $r = \gamma, \theta = \frac{\pi}{2}, s = -\kappa$ in Eq. (10). As a result, real eigenvalues are expected as long as the PT symmetry is not broken. A direct diagonalization gives the two eigenvalues,

$$\beta_\pm = \pm\sqrt{\kappa^2 - \gamma^2}. \tag{19}$$

If the gain/loss parameter is smaller than the coupling coefficient $\gamma \leq \kappa$, the propagation constants $\xi_\pm = k + \beta_\pm$ of the two eigenmodes take real values which correspond to exact or unbroken PT-symmetric phase. If $\gamma > \kappa$, on the other hand, PT symmetry automatically fails and the two eigenvalues (19) become a complex conjugate pair. The transition point $\gamma = \kappa$ is known as the spontaneous PT phase-transition point, or more generally the EP.

The eigenstates involve a left and right biorthogonal set of eigenvectors that are defined as $H|\beta_{\pm_R}\rangle = \beta_\pm|\beta_{\pm_R}\rangle$ and $\langle\beta_{\pm_L}|H = \beta_\pm\langle\beta_{\pm_L}|$, with the orthonormalization condition $\langle\beta_{\pm_L}|\beta_{\pm_R}\rangle = \delta_{+-}$. The corresponding eigenstates can be either calculated directly or from Eq. (13) as

$$|\beta_{+R}\rangle = \frac{1}{\sqrt{2\cos\alpha}}\begin{pmatrix} e^{i\alpha/2} \\ e^{-i\alpha/2} \end{pmatrix}; \quad |\beta_{-R}\rangle = \frac{1}{\sqrt{2\cos\alpha}}\begin{pmatrix} ie^{-i\alpha/2} \\ -ie^{i\alpha/2} \end{pmatrix}, \text{ with } \sin\alpha = \frac{\gamma}{\kappa}. \tag{20}$$

In the unbroken PT phase, both the $H$ and $PT$ operators share the same set of eigenstates and the mode intensity in this regime is symmetric with respect to the mirror axis of the two parts of the system, see Fig. 2(b). In the broken region, however, the eigenstates of $H$ cease to be eigenstates of the $PT$ operator, in spite of the fact that $H$ and $PT$ still commute. This stems from the antilinear property of the $PT$ operator. In addition, the spatial distribution of the modes is asymmetric, one of them living predominantly in the amplifying site and the other vanishing in the lossy one (see Fig. 2(b)).

The beam dynamics associated with Eq. (18) is relatively straightforward and was investigated theoretically in [19,22]. To this end, the Hamiltonian in Eq. (18) is first written in the form $H = \beta_\pm\vec{\sigma}\cdot\hat{n} = \beta_\pm\sigma_n$ where $\sigma_n$ is the Pauli matrix projected along the $\hat{n} = \frac{1}{\beta_\pm}(-\kappa, 0, i\gamma)$ direction. By applying the identity $e^{id\sigma_n} = I\cos d + i\sigma_n \sin d$, the effective evolution operator $U(z)$ takes the form

$$U(z) = e^{-izH} = I\cos(\beta_\pm z) - i\frac{H}{\beta_\pm}\sin(\beta_\pm z). \tag{21}$$

Application of the above operator (21) to a generic initial preparation $\phi(z = 0) = \begin{pmatrix} c_1 \\ c_2 \end{pmatrix}$ permits to evaluate the beam $\psi(z)$ at a propagation distance $z$,

$$\phi(z) \equiv \begin{pmatrix} \phi_1(z) \\ \phi_2(z) \end{pmatrix} = \frac{1}{\cos\alpha} \begin{pmatrix} c_1 \cos(\beta_+ z/2 - \alpha) - ic_2 \sin(\beta_+ z/2) \\ c_2 \cos(\beta_+ z/2 + \alpha) - ic_1 \sin(\beta_+ z/2) \end{pmatrix}. \tag{22}$$

The total light intensity $I(z) = |\phi_1(z)|^2 + |\phi_2(z)|^2$ is not any more a constant of motion. Its dependence on the paraxial distance $z$ can be easily computed from Eq. (22),

$$I(z) = \frac{1}{\cos^2\alpha}\left[c_1^2\cos^2\left(\frac{\beta_+ z}{2} - \alpha\right) + c_2^2\sin^2\left(\frac{\beta_+ z}{2}\right)\right]\left[c_2^2\cos^2\left(\frac{\beta_+ z}{2} + \alpha\right) + c_1^2\sin^2\left(\frac{\beta_+ z}{2}\right)\right]. \tag{23}$$

Figures 2(c)-(e) illustrate the beam dynamics described by Eqs. (22) and (23) in a PT-symmetric coupler for some typical values of the gain/loss parameter $\gamma$. The associated experimental measurements have been performed in [39,40] with use of coupled optical waveguides. In these experiments, the authors recognized that as $\gamma$ reaches $\kappa$, the total beam power starts to grow exponentially, while for $\gamma < \kappa$ (coherent) power oscillations are observed. The former behavior is rooted in the complex nature of the propagating constants in the broken phase, while the latter is due to the biorthogonal nature of the two supermodes. At $\gamma = \kappa$ the intensity grows in a power law manner with respect to the propagation distance $z$, signaling the existence of a defective eigenvalues. All these cases can be easily derived from Eq. (22) analytically. On the other hand, in all cases the beam evolution is nonsymmetric. Specifically, the beam propagation pattern differs depending on whether the initial excitation is on the left or right waveguides. This has to be contrasted with the Hermitian case ($\gamma = 0$), where the beam propagation is insensitive to the initial condition.

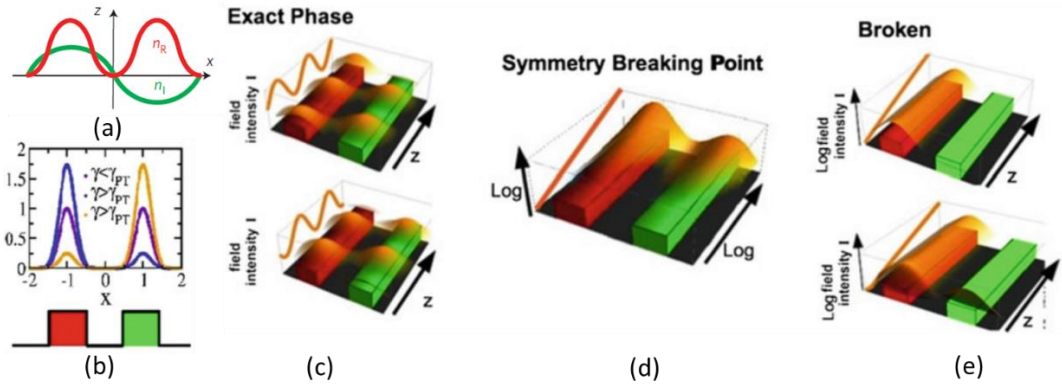

Figure 2 PT-symmetric coupler. (a) Schematic of the optical PT potential for two waveguide coupler. (b) Evolution of typical spatial distribution of the two supermodes. (c)-(e) Beam dynamics for this PT-symmetric coupler. (c) In the unbroken phase region, the total light intensity shows Rabi-like oscillatory behavior and the dynamics is asymmetric for different initial conditions with respect to the z-axis of symmetry of the structure. (d) The beam dynamics at the exceptional point where the total intensity grows in a power law along with the propagation distance z. (e) In the phase breaking regime, the beam intensity grows exponentially in an asymmetric way. ((b)-(e) are adapted from Ref. [22])

Though the analogy enables the possibility of theoretically extending PT symmetry from quantum to wave optics, it is the recognition of coupled-mode dynamics that actually paves the way for experiment involved. In the language of coupled-mode theory, each matrix element in the Hamiltonian (18) has its well-defined physical meaning. This makes itself distinct from the quantum counterpart (10), which more or less suffers the perplex of metaphysics. Thanks to its wide range of applicability, coupled-mode theory has already been shown to exceedingly facilitate the studies of PT symmetry in a variety of branches of physics. To date, all research on PT symmetry in the classical world cannot escape the coupled-mode dynamics. Setting aside our theme, we notice that the terminology "PT symmetry" [177,178] was used to explain some optical experiments with high-temperature superconductors long before the invention of PT symmetric quantum mechanics [33].

3.2 PT symmetry in coupled optical microcavities

It turns out that the formulation of a PT-symmetric coupler in Section 3.1 utilizing the coupled-mode theory is fairly general for various physical systems in terms of scattering problems. In optical settings, light scattering delivers important information about their spectral properties and reveals the consequences of their symmetries. The adoption of PT potentials is expected to empower new regimes of linear scattering, and furthermore facilitate novel photonic devices that surmount to conservative architectures. In the last years silicon photonics has become one of the most promising photonic integration platforms. This can be mainly attributed to the combination of a very high-index contrast and the availability of mature CMOS fabrication technology, which allows the use of electronics fabrication facilities to make photonic circuitry. Thanks to the CMOS technique, silicon waveguide structures have shown unprecedented reduction in footprint, and especially wavelength-selective devices. A prime example of this is optical microring resonators (a set of waveguides in which at least one is a closed loop coupled to some sort of light input and output). Optical microresonators [1-3] offer the potentials for developing new types of photonic devices such as light emitting diodes, low-threshold microlasers, ultra-small optical filters and switches for wavelength-division-multiplexed (WDM) networks, color displays, and so on. Moreover, novel designs of microresonators open up very challenging fundamental science applications beyond optoelectronic device technologies. The interaction of active or reactive material with the modal fields of optical microresonators provides key physical models for basic research including cavity QED experiments, spontaneous emission control, nonlinear optics, bio chemical sensing and even quantum information processing.

In an optical waveguide light propagation is confined via total internal reflection (TIR) by two dielectric interfaces to a region of high refractive index. The guiding region of high refractive index is formally equivalent to a potential wall wherein the electric fields may be decomposed into eigenmodes that are solutions of the Schrödinger equation. Light propagating in curved waveguide (such as cylindrical or spherical surface) is still guided via TIR at the outer interface, but it no longer demands an inner interface to complete the confinement. Elimination of the inner boundary leaves a dielectric disk that supports whispering gallery modes (WGMs). These modes consist of azimuthally propagating fields guided by TIR at the dielectric interface and *optical inertia* that prevents the field from penetrating inward beyond a fixed radius termed the inner caustic. Mathematically, a WGM is a solution of the Helmholtz equation in a curved coordinate geometry. Attention is restricted to a cylindrical geometry appropriate for the analysis of planar disk and ring resonators.

Inspired by waveguides experiments in the spatial domain [39,40], in 2014 two groups independently performed optical PT symmetry using two directly coupled WGM microtoroid resonators in the temporal domain [4,5]. As schematically depicted in Fig. 3, the dynamics of the input signal field in two coupled cavities is effectively described by the temporal coupled-mode approach [176],

$$i\frac{d}{dt}A = HA \text{ with } A = \begin{pmatrix} a_g \\ a_\gamma \end{pmatrix} \text{ and } H = \begin{pmatrix} ig & \mu \\ \mu & -i\gamma \end{pmatrix}. \tag{24}$$

Here, $a_{g,\gamma}(t)$ represent the normalized supermode amplitudes in toroids 1 and 2, and $\mu$ stands for the coupling strength between two resonators due to mode overlapping. $g = \frac{g_0}{1+|a_g/a_s|^2} - \gamma_g$ [4] is the net gain supplied by optically pumping doped $Er^{+3}$ ions in the active toroid 1, where $g_0$ is the pumping gain for $a_g = 0$, $a_s$ associates with the gain saturation threshold, and $\gamma_g$ counts the total loss. Similarly, $\gamma$ characterizes the total loss for the passive cavity 2, which relies on the quality factor $Q$. In the derivation of Eq. (24), the two microcavities are assumed to share the same resonant frequency, which also coincides with the input signal frequency. The nonlinearity appearing in $g$ seemingly makes the problem difficult to analyze in general. In fact, depending on the circulated optical power $|a_g|^2$ with respect to the saturation threshold, it can be segmented into two different regions: linear and nonlinear. That is, if $|a_g/a_s| \ll 1$, the nonlinear saturation effect can be neglected safely. On the other hand, if $|a_g/a_s| \gg 1$ one has to take into account the nonlinearity, which contains rich physics (see Sections 3.5 and 3.6) despite its complexity.

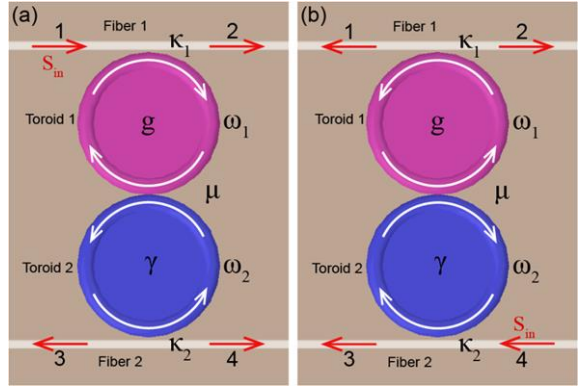

Figure 3 On-chip two directly coupled WGM microtoroid resonators with gain and loss for PT symmetry as well as optical nonreciprocity. The system is composed of two toroids (1 and 2) which are coupled to two tapered optical fibers (1 and 2). Toroid 1 is an active resonator resonant at frequency $\omega_0$ and is optically pumped to produce a net gain $g$. Toroid 2 is a passive resonator resonant also at $\omega_0$ with a decay rate $\gamma$. The coupling strengths between toroid 1-toroid 2, toroid 1-fiber 1, and toroid 2-fiber 2 are, respectively, denoted by $\mu$, $\kappa_1$ and $\kappa_2$. Optical reciprocity and nonreciprocity are studied by analyzing transmittance spectra in forward (a) and backward (b) propagation configurations. $s_{in}$ represents the amplitude of the input signal field.

Let us first focus on the linear case of $|a_g/a_s| \ll 1$. In this region, one can quickly check that for perfect PT symmetry, it demands balanced gain and loss $g = \gamma$ in Eq. (24). As such, the PT Hamiltonian (24) concurs with that (18) for the waveguide directional coupler. Without any difficulty, the two eigenspectral branches are centered at

$$\omega_\pm = \pm\sqrt{\mu^2 - \gamma^2}, \tag{25}$$

pertaining to the cavity resonance frequency. By adopting the same method ascribed in Section 3.1, the light beam $\phi(t)$ at the propagation time $t$, for the initial preparation $A(t=0) = \begin{pmatrix} c_1 \\ c_2 \end{pmatrix}$, takes the form

$$A(t) \equiv \begin{pmatrix} a_g(t) \\ a_\gamma(t) \end{pmatrix} = \frac{1}{\cos\alpha} \begin{pmatrix} c_1 \cos(\omega_+ t/2 - \alpha) - ic_2 \sin(\omega_+ t/2) \\ c_2 \cos(\omega_+ t/2 + \alpha) - ic_1 \sin(\omega_+ t/2) \end{pmatrix} \tag{26}$$

with $\alpha = \arcsin(\gamma/\mu)$.

From the eigenspectra (25), a PT-like spectral bifurcation occurs around the exceptional point $\mu = \gamma$. For $\mu \geq \gamma$, the two PT supermodes with a zero linewidth are spectrally distributed at $\pm\sqrt{\mu^2 - \gamma^2}$ away from the cavity resonance and the optical power does a Rabi-like oscillation between two cavities (see Eq. (26)). By operating in such a region, an ideal single-mode lasing operation can be realized with the spectral width much less than the cavity linewidth, see Section 4.1. As $\mu < \gamma$, $\omega_\pm$ become complex and the PT-symmetry is spontaneously broken. As the PT-symmetric phase is broken, one supermode gradually vanishes because of the absorption while the other experiences amplification. One peculiar feature occurring in this latter case is that while the eigenspectra are complex, their linewidths are considerably shrunk as a result of the compensation from the mode coupling. At $\mu = \gamma$ the two modes start to coalesce into the cavity resonance frequency. Moreover, the presence of this EP leads to the substantial reduction of the lasing threshold [118,119]. Generally, the existence of this singularity point does not require the exact balance of gain and loss between two components, which gives rise to the opportunity to introduce the concept of *passive PT symmetry* [39]. Using two passively-coupled microtoroid resonators, in a recent experiment [179] Yang's group realized loss-induced suppression and revival of lasing, a phenomenon that resembles a temporal version of the loss-induced transparency observed in two passively-coupled waveguides [39]. Although an effective PT system can be mathematically constructed by pulling out a global loss offset in passive structures, in this review we retain our attention to *active* PT symmetry with balanced gain and loss. For passive-PT configurations, unless otherwise specified, we will reserve them as conventionally coupled systems as they have been researched for a very long time from a different perspective.

Although Eq. (26) seems to give a good account of the light propagation within two microresonators, it does not take into account the light launching mechanism properly, which leads to additional features beyond the linear homogeneous systems of ordinary differential equations (24) with constant coefficients. This deviates from the waveguide case and essentially leads to different dynamics. To count the light injection consistently, light transport in such a coupled-cavity system can be studied by the input-output theory [176] which results in inhomogeneous linear systems of ordinary differential equations. To fulfill the PT operation, the output signal amplitudes at ports 4 and 1, corresponding to the forward and backward propagation configurations (Fig. 3) respectively, are of interest for the same input signal amplitude $s_{in}$. That is,

$$\begin{cases} i\frac{da_g}{dt} = i\gamma a_g + \mu a_\gamma + i\sqrt{\kappa_1}s_{in}, \\ i\frac{da_\gamma}{dt} = -i\gamma a_\gamma + \mu a_g, \quad \text{(forward)} \\ s_4^F = \sqrt{\kappa_2}a_\gamma, \end{cases} \qquad (27)$$

and

$$\begin{cases} i\frac{db_g}{dt} = i\gamma b_g + \mu b_\gamma, \\ i\frac{db_\gamma}{dt} = -i\gamma b_\gamma + \mu b_g + i\sqrt{\kappa_2}s_{in}, \quad \text{(backward)} \\ s_1^B = \sqrt{\kappa_1}b_g. \end{cases} \qquad (28)$$

These two sets of equations (27) and (28), apparently different from Eq. (26), are then the starting points for analyzing the PT-symmetric optics (and nonreciprocal signal transmission). The input-output theory does not change the cavity resonance property, but it does modify the supermodes dynamics. Alternatively, the centers of two PT eigenspectra still follow Eq. (25). Yet, the signal transmission can be simply evaluated in the steady-state approximation. As such, from Eqs. (27) and (28), the transmission coefficients at ports 4 and 1 are, respectively,

$$T_4^F(\omega) = \left|\frac{s_4^F}{s_{in}}\right|^2 = \frac{\mu^2\kappa_1\kappa_2}{|\Gamma(\omega)|^2} = T_1^B(\omega) = \left|\frac{s_1^B}{s_{in}}\right|^2 \qquad (29)$$

with $\Gamma(\omega) = \mu^2 - \gamma^2 - (\omega - \omega_0)^2$. Equation (29) immediately leads to the conclusion that the linear PT symmetry yields symmetric (or reciprocal) transmission spectra regardless of the unbroken or broken phase. Additionally, it shows that due to spectral singularities, these transmission coefficients tend to grow infinitely, a phenomenon that was previously noticed by Mostafazadeh [120] because of the linearity of the system. This feature of significant signal amplification offers a practical way to determine whether $g$ balances $\gamma$ in the experiment.

These theoretical analysis have been confirmed in the experiments [4,5] where the experimental setup is schematically shown in Fig. 4(a). Figure 4(b) is the 2D top view of the system. The evolution of the two PT supermodes from the unbroken phase to the broken phase is illustrated in Fig. 4(c) as a function of the separation distance between two microcavities. If one focuses on the peak locations of the two PT eigenspectra, Fig. 4(d) clearly reveals the quadratic splitting between two PT modes as well as the transition from the unbroken phase to the broken. It is worth noting that because of the imperfect cavity surface, backscattering originated from the Rayleigh scattering [180] is non-negligible in high-Q cavities. The observed double-peak structures in Fig. 4(c) are precisely the result of this reason. Owing to the potential application in integrated photonics, the research on PT symmetry using optical microcavities has attracted a considerable attention in the past few years. For example, Phang *et al* theoretically considered the effects in the presence of dispersion and frequency misalignment in two PT-symmetric coupled microresonators [181]. He *et al* further generalized the case to an infinite structure with the cyclic permutation symmetry [92].

Despite the mechanism of PT symmetry breaking is inherently linear, of fundamental interest will be to comprehend how this process unfolds in the presence of nonlinearity [22-26]. Given that lasers are by nature nonlinear devices, this trend becomes imperative. Likewise, adding nonlinearity was predicted to support ample novel phenomena including solitons in PT structures [38]. The appearance of gain saturation in the active optical microcavity [4,5] provides a new addition along this direction. In the presence of saturable gain and loss that is prevalent in such kind of systems, optical properties of PT-symmetrically coupled microcavities would be dramatically modified and may exhibit new features that cannot be achievable in the linear case. By counting the saturation effects in the dual optical micoresonator arrangement shown in Fig. 3, the light transport is now changed to

$$\begin{cases} i\frac{da_g}{dt} = -i\gamma_g a_g + i\frac{g_0}{1+|a_g/a_s|^2} a_g + \mu a_\gamma + i\sqrt{\kappa_1}s_{in}, \\ \quad i\frac{da_\gamma}{dt} = -i\gamma a_\gamma - i\frac{\gamma_0}{1+|a_\gamma/a_s|^2} a_\gamma + \mu a_g, \qquad \text{(forward)} \\ \qquad s_4^F = \sqrt{\kappa_2}a_\gamma, \end{cases} \qquad (30)$$

and

$$\begin{cases} \quad i\frac{db_g}{dt} = -i\gamma_g b_g + i\frac{g_0}{1+|b_g/a_s|^2} b_g + \mu b_\gamma, \\ i\frac{db_\gamma}{dt} = -i\gamma b_\gamma - i\frac{\gamma_0}{1+|b_\gamma/a_s|^2} b_\gamma + \mu b_g + i\sqrt{\kappa_2}s_{in}, \quad \text{(backward)} \\ \qquad s_1^B = \sqrt{\kappa_1}b_g. \end{cases} \qquad (31)$$

respectively, for the forward and backward configurations, where $\gamma_0$ represents the unsaturated loss for the passive cavity. Here it assumes the same threshold $a_s$ for both gain and loss microresonators and $\gamma_g = \gamma$. The saturable loss term could be introduced through an externally controllable probe to induce in practice. When the modal field amplitudes are small in both injection cases, Eqs. (30) and (31) are reduced to the linear situations (27) and (28) by ignoring the saturation effects. Caution should be paid to the coupling strengths $\kappa_1$ and $\kappa_2$. Depending on $|\sqrt{\kappa_1}s_{in}| \gtrsim |g_0 a_s|$ and/or $|\sqrt{\kappa_2}s_{in}| \gtrsim |\gamma_0 a_s|$ as well as $|a_{g,\gamma}| \gtrsim |a_s|$ and/or $|b_{g,\gamma}| \gtrsim |a_s|$, the modal field amplitudes in the system can experience very complicated evolutions and result in rich physics including the mixtures and transitions of linear and nonlinear propagations even for the same propagation configuration. Due to the nonlinearity, Eqs. (30) and (31) are difficult to be solved analytically in general. An interesting attempt was recently made by Hassan and his coworkers [182], which will not be further discussed in this article.

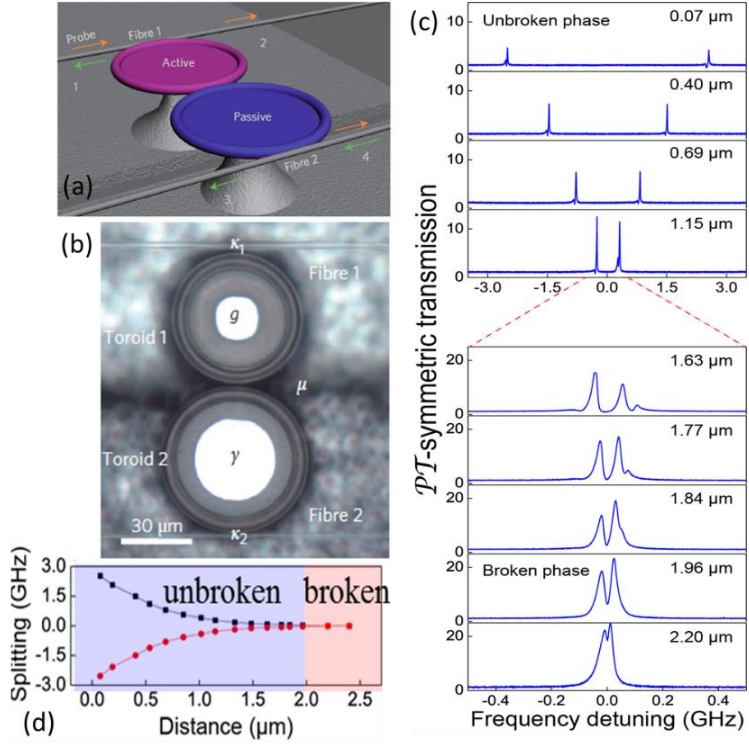

Figure 4 PT symmetry using two directly coupled WGM microtoroid resonators with balanced gain and loss. (a) Schematic 3D view of the system. (b) Top view optical microscope image of the system in (a). (c) Evolution of PT-symmetric transmission spectra from the unbroken phase to the broken by varying the coupling strength (or separation distance) of the two toroids. (d) Spectral splitting between the two PT supermodes in (c). (Adapted from Ref. [4])

### 3.3 Proposal on anti-PT symmetry in optical microcavities

Though PT symmetry in various physical settings has been studied extensively in the past decades, its counterpart, anti-PT symmetry, has only been known by people in very recent years. One major challenge stems from the unusual dissipative coupling appearing in the off-diagonal terms of the anti-PT-symmetric Hamiltonian (14). As it will become clear shortly, such a dissipative coupling mechanism is fundamentally different from the Hermitian couplings in any PT-symmetric Hamiltonian. By employing atomic coherence transport in a warm alkali atomic vapor cell, the first anti-PT experiment [64] was demonstrated in two coupled atomic spin waves, where the randomness and irreversibility associated with the atomic thermal motion set the foundation of non-Hermiticity of the system Hamiltonian. As mentioned in Section 2.3, the dynamics of an anti-PT system is governed by a non-Hermitian Hamiltonian anti-commuting with the joint anti-linear PT operator, or $\{PT, H_{\text{anti}}\} = 0$. Moreover, the mathematical transformation between the PT-symmetric and anti-PT-symmetric Hamiltonians suggests that the properties of an anti-PT system are exactly conjugate to the corresponding PT system. For instance, in the symmetry-unbroken phase, lossless propagation in a PT system corresponds to refractionless (i.e. unit-refraction) propagation in an anti-PT system [64]. This rather intriguing effect gives rise to a complementary probe in the non-Hermitian optics, stretches the extent of novel techniques for light manipulation, and offers new paradigms of light control. In spite of these attractions, anti-PT symmetry is not easily attainable in experiment. In two earlier attempts, one proposal [183] requires a challenging balance of positive and negative refractive indices in coupled metamaterials, while the other [184] utilizes an optical lattice of spatially driven cold atoms, which is very difficult in practice and impossible to scale.

Motivated by real applications, an implementation of anti-PT symmetry in an integrable photonic system becomes thus highly desirable. It turns out that anti-PT symmetry is realizable in three passively-coupled optical microcavities [185,186] through the technique of adiabatic elimination [187]. As schematically depicted in Fig. 5(a), $\omega_j, \Delta_j,$ and $\gamma_j$ ($j = 1,2,3$) denote, respectively, the cavity resonance frequency, laser frequency detuning, and the total loss rate in microcavity $j$. $\mu_{1(2)}$ and $\kappa_{1(2)}$ stand for the coupling strengths between Cavity 2 and Cavity 1 (3), and in/out of Cavity 1 (3) through tapered fiber, respectively. Note that there is no direct coupling between Cavity 1 and Cavity 3. For the forward configuration, the signal laser is launched from Cavity 1 whereas for the backward configuration, the signal field is launched from Cavity 3. The signal dynamics in the system can then be well described by the temporal coupled-mode formalism combined with the input-output theory. For simplicity, let us first look at the signal transport in the forward configuration, which obey the following set of ordinary differential equations,

$$\dot{a}_1 = (i\Delta_1 - \gamma_1)a_1 - i\mu_1 a_2 + \sqrt{\kappa_1} s_{in}, \tag{32}$$
$$\dot{a}_2 = (i\Delta_2 - \gamma_2)a_2 - i\mu_1 a_1 - i\mu_2 a_3, \tag{33}$$
$$\dot{a}_3 = (i\Delta_3 - \gamma_3)a_3 - i\mu_2 a_2. \tag{34}$$

The backward signal transmission can be similarly derived, except that the driving term shall be added to Cavity 3 instead of Cavity 1,
$$\dot{b}_1 = (i\Delta_1 - \gamma_1)b_1 - i\mu_1 b_2,$$
$$\dot{b}_2 = (i\Delta_2 - \gamma_2)b_2 - i\mu_1 b_1 - i\mu_2 b_3,$$
$$\dot{b}_3 = (i\Delta_3 - \gamma_3)b_3 - i\mu_2 b_2 + \sqrt{\kappa_2} s_{in}.$$

To realize anti-PT symmetry, it essentially requires to have the dissipative coupling between optical modes in Cavity 1 and Cavity 3 by adiabatically eliminating optical modes in Cavity 2. As such, Cavity 2 is assumed to be highly lossy in comparison with its laser frequency detuning, i.e., $\Delta_2 \ll \gamma_2$. Under this condition, optical modes in Cavity 1 and Cavity 3 will then become anti-PT symmetrically coupled. That is, from Eq. (33) one has $a_2 \approx -i\frac{\mu_1 a_1 + \mu_2 a_3}{\gamma_2}$. Substituting this result into the rest two equations (32) and (34) yields

$$i\frac{d}{dt}\begin{pmatrix} a_1 \\ a_3 \end{pmatrix} = H_{\text{anti-PT}}\begin{pmatrix} a_1 \\ a_3 \end{pmatrix} + \begin{pmatrix} \sqrt{\kappa_1} s_{in} \\ 0 \end{pmatrix} \text{ and } H_{\text{anti-PT}} = \begin{bmatrix} \Delta_1 + i\left(\frac{\mu_1^2}{\gamma_2} + \gamma_1\right) & i\frac{\mu_1 \mu_2}{\gamma_2} \\ i\frac{\mu_1 \mu_2}{\gamma_2} & \Delta_3 + i\left(\frac{\mu_2^2}{\gamma_2} + \gamma_3\right) \end{bmatrix}. \tag{35}$$

The physics now becomes clear. In order to make the non-Hermitian Hamiltonian (35) satisfy $\{PT, H_{\text{anti-PT}}\} = 0$, besides $\mu_1 = \mu_2$ and $\gamma_1 = \gamma_3$, the detunings $\Delta_1$ and $\Delta_3$ should be equally but oppositely tuned in experiment, same as reported in Ref. [64]. As a result, the Hamiltonian in Eq. (35) coincides with the one given in (14). The two off-diagonal coupling coefficients are indeed purely imaginary as demanded in Eq. (14). In addition, these two non-Hermitian terms, in fact, violate the time-reversal symmetry, thus implying nonreciprocal light transmission. It is anticipated that such a linear system may be beneficial to chip-based optical nonreciprocity beyond the conventional Faraday magneto-optical effect. Without any computational difficulty, the two eigenspectra have the form of

$$\varpi_{\pm} = i\left(\frac{\mu_1^2}{\gamma_2} + \gamma_1\right) \pm i\sqrt{\left(\frac{\mu_1^2}{\gamma_2}\right)^2 - \Delta_1^2}. \tag{36}$$

Depending on the signal detuning $\Delta_1$ greater or smaller than the dissipative coupling strength $\frac{\mu_1^2}{\gamma_2}$, the two eigenvalues (36) are either complex or purely imaginary. For complex eigenspectra, their real parts describe the signal transmission peak locations while the same imaginary parts indicate the same spectral linewidth.

The system is in the anti-PT phase breaking regime. On the contrary, in the anti-PT unbroken phase region, purely imaginary eigenspectra mean that the two signal transmission peaks coalesce but possess different linewidths. The anti-PT phase transition occurs at the exceptional point, $\Delta_1 = \frac{\mu_1^2}{\gamma_2}$. These spectral behaviors have been theoretically plotted in Fig. 5(b). In anti-PT symmetry, one interesting effect appearing in the phase unbroken case is to create flat-band structure [64]. The occurrence of the phenomenon relies on the destructive interference associated with different effective refractive indices created in two cavities 1 and 3. It is this destructive interference that leads to the flat-band structure for the signal transmission, whose bandwidth is mainly determined by the anti-PT bandwidth in the unbroken phase region. The imperfections in real implementations may cause the shift of exceptional point and bring Hermitian components into the anti-Hermitian coupling strengths.

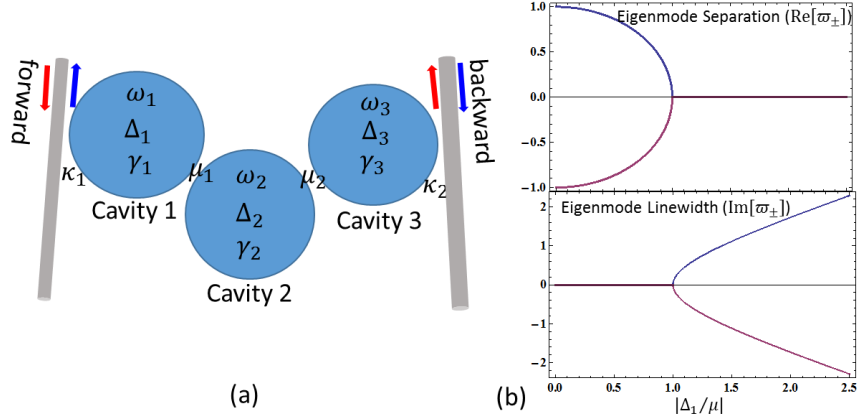

Figure 5 (a) Schematic of the proposed tri-microcavity system for the realization of both PT and anti-PT symmetry. (b) The real (top) and imaginary (bottom) parts of the two supermodes (34) obtained from the transmission spectra.

Interestingly, the proposed structure also allows the implementation of PT symmetry [4,5] in the opposite regime. In this case, it necessitates Cavity 2 to be of high quality in terms of its laser frequency detuning, i.e., $\Delta_2 \gg \gamma_2$. Under this condition, Eq. (33) gives $a_2 \approx \frac{\mu_1 a_1 + \mu_2 a_3}{\Delta_2}$. Plugging this result into Eqs. (32) and (34) then yields

$$i\frac{d}{dt}\begin{pmatrix} a_1 \\ a_3 \end{pmatrix} = H_{\text{PT}} \begin{pmatrix} a_1 \\ a_3 \end{pmatrix} + \begin{pmatrix} \sqrt{\kappa_1} s_{in} \\ 0 \end{pmatrix}, \quad H_{\text{PT}} = \begin{bmatrix} \left(\Delta_1 - \frac{\mu_1^2}{\Delta_2}\right) + i\gamma_1 & -\frac{\mu_1 \mu_2}{\Delta_2} \\ -\frac{\mu_1 \mu_2}{\Delta_2} & \left(\Delta_3 - \frac{\mu_2^2}{\Delta_2}\right) + i\gamma_3 \end{bmatrix}. \quad (37)$$

By comparing the Hamiltonian in Eq. (37) with that in Eq. (27), for perfect PT symmetry it further needs $\gamma_1 = -\gamma_3$ in addition to $\Delta_1 = \Delta_3$ and $\mu_1 = \mu_2$. We notice that the proposed tri-microcavity system is the first architecture capable of realizing both PT and anti-PT.

Owing to their linearity, a different perspective about the relationship of perfect PT and anti-PT symmetry can be taken by recasting Eqs. (27) and (35) into the form of a damped harmonic oscillator driven by a constant force,

$$\ddot{a}_g + (\gamma - g)\dot{a}_g + (\mu^2 - g\gamma)a_g = \xi s_{in}, \quad (38)$$

and

$$\ddot{a}_1 + i(\Delta_1 - \Delta_3)\dot{a}_1 - (\eta^2 + \Delta_1 \Delta_3)a_1 = \zeta s_{in}. \quad (39)$$

Note that in the derivation of Eq. (39), the decays have been set to be zero for simplicity. From Eqs. (38) and (39), real eigenspectra for perfect PT and anti-PT symmetry correspond to have a simple harmonic oscillation by cancelling the damping term.

   3.4 On-chip optical nonreciprocity

Achieving rapid progress in integrated photonic circuits demands all-optical elements for high-speed processing of light signals. The optical isolator is one such indispensable ingredient. Similar to electronic diodes, an optical isolator shall ensure the flow of light to be unidirectional and reduce problems caused by unwanted reflections or spurious interference effects. Owing to the time-reversal symmetry retained in light-matter interaction, unfortunately, light wave transport in a linear, time-invariant optical system complies with the Lorentz reciprocity [188]. As a result, the successful design of an optical isolator relies on the breach of time-reversal symmetry, as typically achievable by applying the Faraday effect in magneto-optical media through the inclusion of anti-symmetric off-diagonal dielectric tensor elements. In spite of their commercial success, this well-established approach poses a severe challenge to implement in chip-scale photonics due to fabrication complexity with the mature CMOS technique, difficulty in locally confining magnetic fields, and significant material losses. As such, a vibrant search for different physical principles to achieve on-chip optical nonreciprocity has garnered a vast impetus in recent years. Alternative methods, most, as yet, far from practical realizations, include indirect interband transitions, optomechanical interactions, Kerr nonlinearities, gain/absorption saturation, thermo-optic effect, opto-acoustic interaction, Raman amplification, nonlinear parametric amplification, stimulated Brillouin scattering, Bragg scattering, and nonlinear nonadiabatic quantum jumps. The progress with the two separate topics of chip-scale nonmagnetic optical isolators and PT-symmetric optics brings forward a legitimate question whether it would be possible to attain optical nonreciprocity with PT symmetry. This has motivated continued interest in the realization of chip-based optical nonreciprocity with PT symmetry.

In the works [4,5], two groups also observed highly nonreciprocal transmission in two gain-loss-coupled microtoroids. As pointed out by Ref. [4], this nonreciprocal light transmission is due to the gain saturation nonlinearity, see Fig. 6(a). Thanks to the high Q factor, it turns out that the signal gain is very easily saturated when increasing the input signal and/or pump powers. With this compound system, they achieved switchable isolation by elevating the incident signal power from a few nanowatts to ~30 µW. Such a switchable isolation behavior can be also obtained, for example, by only modifying the coupling strength $\kappa_2$ between fiber 2 and toroid 2 (Fig. 4(b)) but with fixed input signal/pump powers and other coupling strengths ($\mu, \kappa_1$), as shown in Fig. 6(b). As one can see, when the coupling between fiber 2 and toroid 2 is weak, the isolation ratio takes a negative sign, which implies more backward transmission than forward. However, when the coupling is large enough, the isolation ratio becomes positive, indicating more output from the forward configuration. In the coupled-mode theory (27) and (28) presented in Section 3.2, the theoretical simulations can explain the observed phenomena well by including the gain-saturation nonlinearity. The use of gain-saturation nonlinearity to break the time-reversal symmetry has been further studied subsequently by utilizing only one active microcavity [189,190]. As these latter works have no relation with PT symmetry, we will not discuss them here.

In order to create asymmetric transmission, in the work [81] by Nazari *et al*, they theoretically explored directional nonlinear Fano resonances, which emerge in a photonic circuit consisting of two nonlinear PT-symmetric microcavities which are side-coupled to a waveguide (see the insets in Fig. 7(a)). Due to the interplay of Kerr-nonlinearity with the active elements, the generated nonlinear Fano resonances are triggered at different resonance frequencies depending on the direction of the incident light. Meantime, the PT-symmetric configuration provides a strong output signal and ensures the stability of the circuit against lasing action. By operating the system in the "weak" indirect coupling between two microcavities, the authors numerically showed that the interplay of the stability (i.e. nonlasing) and PT amplification properties in the broken phase together with the presence of Kerr nonlinearity can lead to unidirectional amplified transport (Figs. 7(a) and (b)).

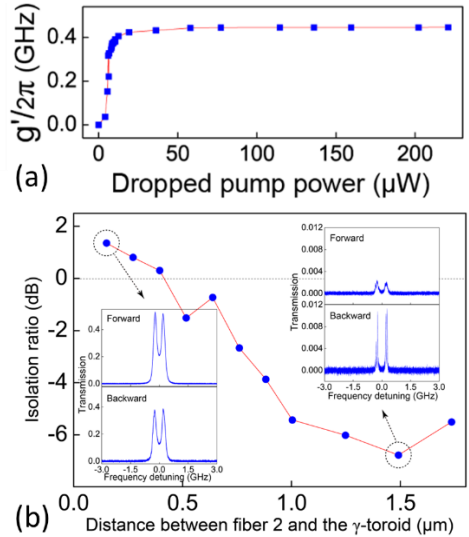

Figure 6 (a) Measured gain $g'$ as a function of the dropped pump power, which clearly reveals the gain-saturation nonlinearity. (b) Observed switchable optical isolation in terms of the separation distance between toroid 2 and fiber 2. The insets are typically output transmission spectra for the forward and back signal inputs. (Adapted from Ref. [4])

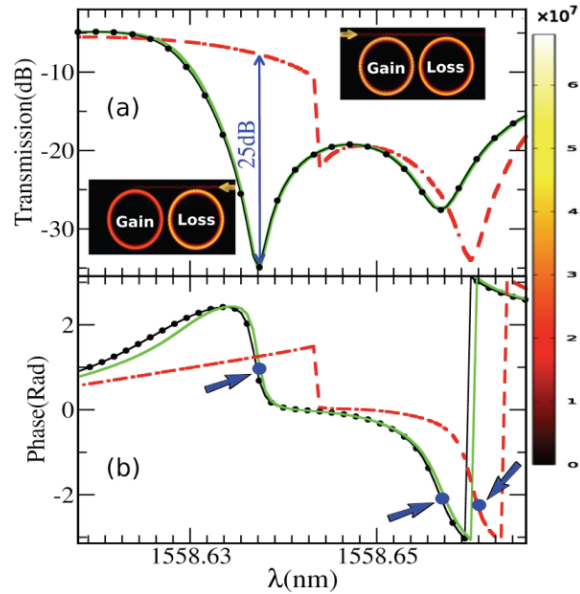

Figure 7 Theoretical proposal on optical isolation via PT-symmetric nonlinear Fano resonances. (a) The transmissions of the linear system for input waves from the gain (filled black circles) and the loss (black line) side are compared with the corresponding transmissions in the case of microcavities with Kerr nonlinearity. The resonance from the lossy side (green line) experiences a small red-shift with respect to the linear structure. In contrast, the transmittance curve (both line-shape and resonance position) of an incident wave entering the structure from the gain side (red dotted-dashed line) is different. Insets: Schematic of the nonlinear PT-symmetric photonic circuit. The color coding indicates the intensity strength of the field inside the microcavity. (b) The transmission phases (in the interval $[-\pi, \pi]$) are plotted as a function of wavelength $\lambda$. The colors and the line-type indicate the same scattering process as the one used

in the upper panel. The blue arrows (and the blue dots) mark the wavelengths where the Fano resonances appear. (Adapted from Ref. [81])

Of these schemes [4,5,81], asymmetric optical transmission is mostly demonstrated with only injecting a light wave in either forward or backward direction but never both. This type of implementation drawbacks was questioned in a recent theoretical proposal by Fan group [191], where they in particular proved that Kerr or Kerr-type nonlinearities are incapable of providing complete isolation because of dynamic reciprocity. A close examination on these schemes [4,5,81] shows that the involved nonlinearities are Kerr or Kerr-type, rendering them subject to the inevitable dynamic reciprocity. As a result, complete isolation of backscattered signals looks out of reach with these schemes. Very lately, Ma et al has successfully attained nonreciprocal PT symmetry by employing the stimulated Brillouin scattering (SBS) in two microtoroid resonators, where the SBS is exploited to produce direction-sensitive gain in the Brillouin cavity [192]. Unrestricted with the dynamic reciprocity, this linear system has shown a remarkable isolation performance over the existing demonstrations [193].

4. Applications of PT-symmetric optical microcavities

In so far, intriguing applications enabled by PT symmetry in optical microcavities have been experimentally demonstrated including microlaser cavities and supersensitive sensing, which will be briefly reviewed in this section.

### 4.1 Single-mode microring lasers

Since lasers are naturally engaged in gain and loss, they provide perhaps the most straightforward stages where PT symmetry finds its important applications to effectively achieve lasing operation with desired spectral and spatial properties. Laser cavities typically support a large number of closely spaced modes because of their large dimensions over one light wavelength. As a result, the outputs from such lasers are subject to random fluctuations and instabilities due to the mode competition for limited gain. Recently, it was experimentally demonstrated that by taking advantage of PT symmetry, new single-mode lasers can be elegantly realized with enhanced single-mode operations and greater tunability [6,7]. The essential idea is to operate the system with a partial PT symmetry, where almost all of the modes in the laser cavities remain in the PT-symmetric phase, except for a single lasing mode which experiences only amplification in the PT-broken phase. Utilizing this idea, two groups [6,7] simultaneously reported the realization of single-mode microring lasers with different configurations.

In the experiment by Hodaei *et al* [6], the single-mode laser was illustrated with two coupled microring resonators, one with gain and one with loss, see Fig. 8. In this setup (similar to Fig. 4), the threshold of PT symmetry breaking depends solely on the relation between gain/loss and the coupling strength, i.e., $g = \mu$. This stringent condition clearly indicates that the threshold for PT symmetry breaking signifies the boundary between amplifying/attenuation and bounded neutral oscillations. That is, any pair of modes, whose gain/loss stays below the coupling strength, will remain neutral. However, as soon as the gain/loss exceeds the coupling, PT symmetry will be broken and a conjugate pair of lasing/decaying modes emerges. The experimental results are summarized in Fig. 8. As a comparison, when there was only one active resonator, or both resonators were active, a similar multimode lasing spectrum ought to be observed in the experiment (see Figs. 8(a) and (b)). In contrast, when the two microrings were active-passive-coupled, a single sharp spectral peak was dominantly left, evidencing the single-mode lasing feature (see Fig. 8(c)).

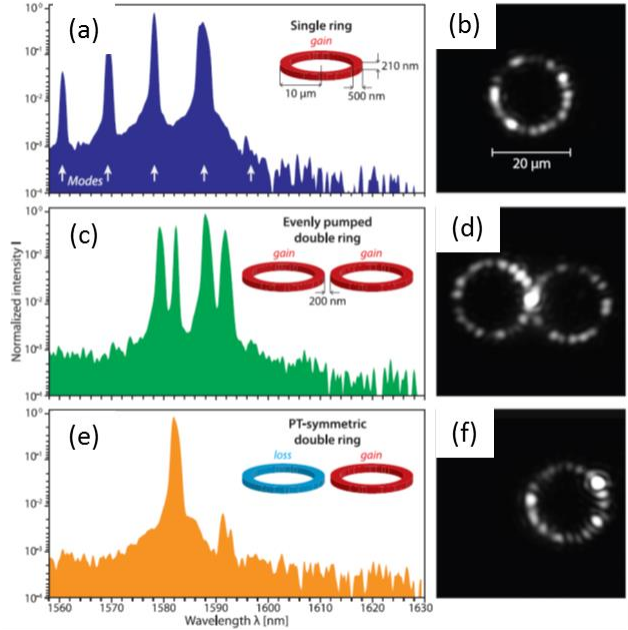

Figure 8 (a) Emission spectrum for one active microring resonator. (b) The corresponding intensity pattern within the ring as observed from scattered light. (c) Spectrum obtained from an evenly pumped pair of the same rings as (a). (d) The intensity pattern shows that both rings equally contribute. (e) Single-mode spectrum under PT-symmetric conditions with gain-loss-coupled rings. (f) Lasing exclusively occurs in the active resonator. (Adapted from Ref. [6])

In the other experiment by Feng et al [7], the single-mode lasing was demonstrated in a microring cavity modulated by a PT-symmetric grating. Different from the former experiment, in this latter one the lasing mode arises as a result of spatial selection instead of spectral selection. Here, the PT-synthetic microring resonator was designed with 500 nm thick InGaAsP multiple quantum wells on an InP substrate (Fig. 9(a)). The gain/loss modulation, satisfying the exact PT-symmetry operation, is periodically introduced using additional Cr/Ge structures on top of the InGaAsP multiple quantum wells along the azimuthal direction. The PT modulation is designed using bilayers on top of the gain material that introduces loss and exactly reverses the sign of the imaginary part of the local modal index, while remaining the same real part. Owing to the continuous rotational symmetry of the microring, PT symmetry breaking here is associated with a thresholdless feature, even if the strength of the gain-loss modulation is infinitesimal. Without the Cr/Ge modulation, the system typically gave a multimode lasing spectrum with different WGM azimuthal orders as shown in Fig. 9(b). However, with the PT index modulation, a pronounced single mode was obtained (Fig. 9(c)). The peak location of this single mode as well as its amplitude remains almost unchanged in comparison with the unmodulated case (Fig. 9(b)).

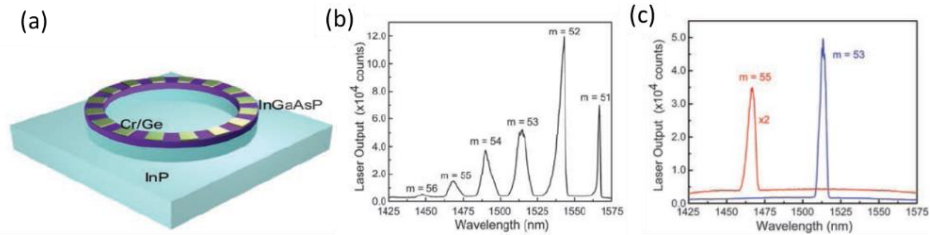

Figure 9 (a) Schematic of the PT microring laser consists of Cr/Ge bilayer structures arranged periodically in the azimuthal direction on top of the InGaAsP/InP microring resonator to mimic a gain-loss modulation. (b) Multimode lasing spectrum typically observed from the microring WGM laser where a number of lasing modes exhibit corresponding to different azimuthal orders. (c) Single-mode lasing spectra obtained from the PT microring shown in (a) by operating at the m = 53 and m = 55 azimuthal orders. (Adapted from Ref. [7])

By overcoming the limited nanocontrol over the phase of light inside the microring resonator, Miao et al [194] recently further realized single-mode orbital angular momentum microlaser on an InGaAsP/InP platform with on-top Ge and Cr/Ge modulations. Different sidewall scatters of the ring coherently radiate light of different phases that continuously vary from 0 to $2\pi$ over one circle along the azimuthal direction. These PT works provide intuitive examples for reassessing the role of optical losses presented in laser systems. Moreover, the peculiar behavior of complex eigenvalues around the exceptional point manifests itself in the threshold and spectral response of any lasing system. For instance, the exceptional points can lead to an effect of pump-induced suppression and revival of lasing, as experimentally demonstrated in coupled quantum cascaded lasers as well as in a pair of silica microcavities with Raman gain [195,196].

4.2 Coherent perfect laser-absorber

Optical gain materials allow light amplification of stimulated electromagnetic radiation in cavities to conquer the undesirable absorption and thus give rise to the birth of lasers. The laser distinguishes itself from other light sources by its inherent coherence in both space and time. Interestingly, the time-reversed counterpart of laser emission, named coherent perfect absorber (CPA) [197], shows that incident coherent optical fields can be perfectly absorbed by a time-reversed optical cavity, where the gain is replaced with an equal amount of loss. Also, the incident fields and frequency should coincide with those of corresponding lasing modes with gain under time symmetry. CPA essentially exploits destructive interference to suppress scattering in a photonic system so that light is completely trapped in a pre-specified spatial region for perfect absorption (see Figs. 10(a) and (b)). The first CPA (or anti-laser) experiment was demonstrated by Wan *et al* [198], based on interferometric control of absorption to perfectly annihilate both the transmitted and reflected light waves.

It seems that laser action and CPA might suppress each other as they are the respective time reversal bodies. Counterintuitively, the CPA-laser concept was first proposed theoretically in a PT system with homogeneous gain and loss embedded under a uniform index grating by Longhi in 2010 [118]. Because of the time-reversal property, the lasing and CPA modes are expected to share common resonant properties including identical frequency dependence, coherent in-phase response and fine spectral resolution. These predictions were confirmed experimentally by Wong *et al* [8] using a straight waveguide of 500 nm thick InGaAsP multiple quantum wells as a gain medium on an InP substrate. The alternating PT-symmetric gain-loss modulation was introduced by periodically placing thin absorbing Cr/Ge structures on top of the waveguide (see Fig. 10(c)), similar as the design for PT-symmetric single-mode microring laser. With such a configuration, the lasing and anti-lasing eigenmodes are degenerate at the boundary of the Brillouin zone owing to the distributed Bragg feedback through the periodic gain-loss modulation. By carefully devising the optical transfer matrix of the two-port system using the coupled-mode theory, the CPA-laser can uniquely satisfy both the lasing and anti-lasing conditions, and approach the CPA-laser point. For example, when the incoming probe beams are $-\pi/2$ offset in phase, the Bragg resonant electric fields constructively interfere only in the gain regions, which leads to strongly amplified outgoing waves and a sharp peak in the output spectrum corresponding to the lasing mode as illustrated in Fig. 10(a). In contrast, when the incoming probe beams are $\pi/2$ offset in phase, the device falls in the anti-lasing mode, where the Bragg interference of electric fields is strongly confined only in the loss regions, causing strong absorption and a narrow dip in the output spectrum as shown in Fig. 10(b). Their main results are summarized in Fig. 10(d). Apparently, at 1,555.8 nm, with the $-\pi/2$ phase-offset probe beams, the maximum Θ (red), manifests a distinct amplification peak of 15 dB, corresponding to the resonant lasing mode; while with the $\pi/2$ phase-offset probe beams, the minimum Θ (blue), is associated with the anti-lasing mode inducing an absorption dip down to –15 dB. This essentially confirms that coherent amplification and absorption is achieved in a single lasing cavity. The lasing and anti-lasing modes share similar resonant wavelengths, and have symmetric amplification and absorption magnitude, all due to the complex conjugate effective indices supported by the PT -broken phase.

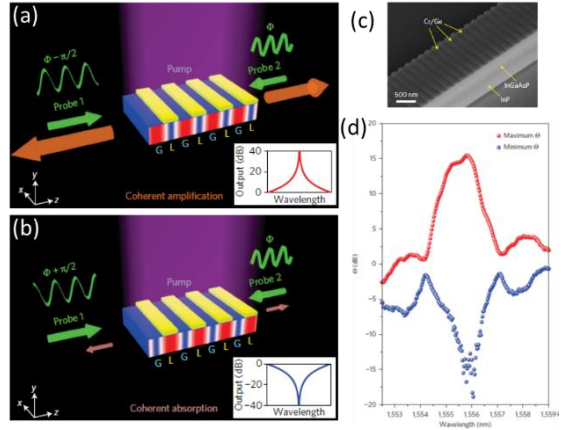

Figure 10  (a) and (b) Schematics showing the principle of the CPA–laser, where the pure gain (G)–loss (L) PT modulation is introduced by placing periodic loss structures on a semiconductor InGaAsP/InP gain waveguide. The PT modulation period is half of the effective wavelength of guided light, to meet the Bragg resonant condition. As such, by uniformly pumping the device, both lasing and anti-lasing eigenmodes can be attained within the same cavity and be selectively excited with the coherent interferometric phase control on the guided light incoming from both directions. (c) SEM image of the fabricated CPA–laser on the InGaAsP/InP platform. (d) The spectra output coefficient Θ of the CPA–laser at the lasing threshold. (Adapted from Ref. [8])

The demonstration provides another effective route of light manipulation and control through the interplay between material loss and gain by PT symmetry. The switchable lasing and anti-lasing modes offer alternative coherent control strategy with an appreciable amplification-to-absorption contrast. By considering nonlinearity effects (including saturation) associated with lasing, an investigation on incorporating optical nonlinearity may be worthwhile for further theoretical and experimental studies.

    4.3 Supersensitive sensing

It is known that a degeneracy of resonant frequencies can serve as a basic element of a sensor since a small perturbation can lift the degeneracy and can thus result in a detectable splitting of these frequencies. This simple principle is usually adopted in designing modern sensor devices including microcavity sensors for single or few particle detection and optical gyroscopes. For Hermitian systems, a perturbation of strength $\delta$ acting on the twofold degeneracy (or the diabolic point (DP)) leads to energy shifts and splitting proportional to $\delta$ (Fig. 11(a)). On the contrary, for an open system, the behavior at a non-Hermitian degeneracy (or EP) is more drastic behaviors than at a DP as at this point in parameter space not only the eigenvalues but also the corresponding eigenvalues coalesce (see Section 2.3). If an EP for two coalescing levels is subject to the perturbation of same strength $\delta$ then the resulting energy splitting is typically proportional to $\sqrt{\delta}$ (Fig. 11(b)). Alternatively, for sufficiently small $\delta$ the splitting is enhanced if compared to a DP even though exactly the same perturbation is applied. In general, for an open system with an Nth-order EP, at which N eigenvalues and the corresponding eigenvectors coalesce, the splitting induced by the perturbation can be proved to scale as $\sqrt[N]{\delta}$. Therefore, for a sufficiently small perturbation $\delta$, the splitting at the EP turns out to be larger. It is precisely this basic characteristics of EPs that was first put forward by Wiersig for supersensitive sensor applications [199,200].

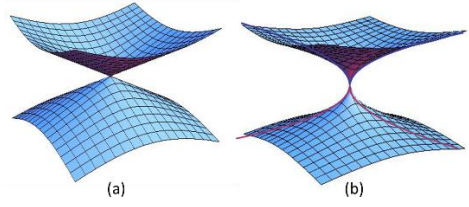

Figure 11 Topology of the surface dictating the complex frequency of diabolic- (a) and exceptional-point (b) sensors, which intuitively shows an energy splitting of the sensors subject to a perturbation δ. For small δ, this energy splitting is proportional to $\sqrt{\delta}$ for EP sensors while δ for DP sensors.

Very recently, this elegant idea has been experimentally tested by two groups using optical microcavities in different configurations: a microtorid cavity with clockwise- and anticlockwise-travelling modes [201], and a ternary PT-symmetric microring system with loss, gain and neutral resonators [9]. In the former, the square –root-perturbation sensitivity was demonstrated in a lossy microcavity by using two silica nano-tips as Rayleigh scatters within the mode volume to tune the coupling between clockwise- and anticlockwise-travelling modes to steer the system to an EP. While in the latter experiment, the higher-order PT EPs were utilized to display the enhanced sensitivity, a cube-root dependence on induced perturbations in the refractive index. Since EPs are generally present in any open system, a system that is non-Hermitian but PT-symmetric is of our primary focus in this article. Hence, in the following we will concentrate on this specific application of PT-symmetrically coupled microrings.

As schematically illustrated in Figs. 12(a1)-(a3), their experimental setup consists of two different PT configurations. Figure 12(a1) depicts such two PT photonic molecules. The first involves two identical cavities, one experiencing gain and the other an equal amount of loss. The second is composed of three resonators with equidistant separations: the two side ring resonators are subjected to equal amounts of gain and loss while the middle ring remains neutral (see Fig. 12(a2)). In addition, the rings evenly exchange energy with each other with a coupling strength $\kappa$. Figure 12(a3) is the SEM image of the fabricated structure. It can be easily shown that the former supports a second-order EP, and the latter a third-order one. As a result, in the first case, the eigenvalues are expected to diverge according to $\sqrt{\delta}$, whereas for the second case, the splitting would be more abrupt because it follows $\sqrt[3]{\delta}$. These features have been indeed verified in their experiments as given in Figs. 12(b1)-(c2). Specifically, in the binary system, once a small frequency mismatch δ is thermally introduced to the optical oscillator around the second-order EP, the two lasing frequencies split according to $\Delta\omega_{EP2} = \sqrt{2\delta\kappa}$. As theoretically expected, Fig. 12(b1) clearly exhibits a square-root wavelength splitting in response to changes in the power dissipated in the heater. The observed linear slope of ½ in the corresponding logarithmic plot affirms this behavior in the inset. Figure 12(b2) gives the measured enhancement in sensitivity in terms of the induced perturbation (i.e. the shift in resonance frequency). As one can see, the enhancement factor increases for small δ. And in this case, the enhancement up to 13 times in the detuning range below 10 GHz is obtained. The sensitivity of the PT tri-microring system was investigated by operating close to one third-order EP. To establish the PT symmetry, the pump light was completely withheld from one of the side rings using a knife edge. Moreover, the central ring is partially illuminated but the third ring is fully pumped. By adjusting the position of the knife edge and the pump power, the three lasing modes of the structure gradually merge into one sharp line, which signals the emergence of a third-order EP. Once this third-order EP was identified, the heater underneath the pumped cavity was activated to introduce the perturbation, which consequently causes the splitting of the single lasing mode into three distinct branches with approximated spectral separation $\Delta\omega_{EP3} \approx 3\sqrt[3]{\delta\kappa^2}/2$. Such a cube-root behavior of the frequency separation $\Delta\omega_{EP3}$ was verified in Fig. 12(c1) as a function of δ. By plotting these experimental data on a logarithmic scale, one can directly infer a slope of 1/3 from the inset. The sensitivity enhancement factor was plotted in Fig. 12(c2), where the sensitivity is magnified about 23 times when the detuning between the gain and neutral rings is below 5 GHz.

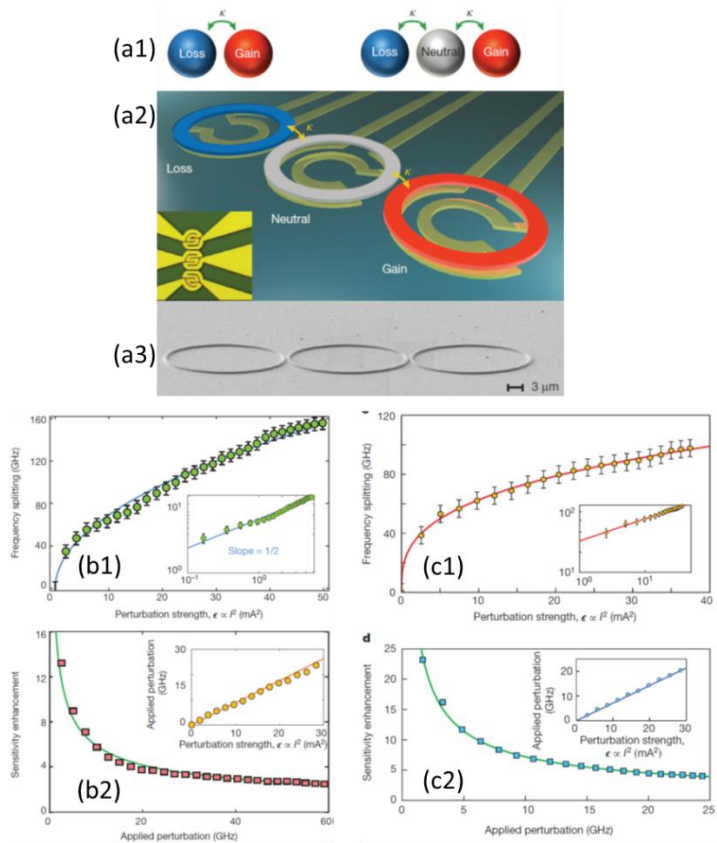

Figure 12 (a1) Schematics of binary (left) and ternary (right) PT-symmetric photonic molecules formed with gain, loss and neutral resonators. (a2) The PT-symmetric tri-microring system with equidistantly spaced cavities. (a3) The SEM image of the structure. Second-order EP improved sensitivity in the binary PT case by observed frequency splitting (b1) and measured enhancement factor (b2). Similarly, third-order EP boosted sensitivity in the ternary PT case by measuring the frequency splitting (c1) and enhancement factor (c2). (Adapted from Ref. [9])

In comparison with the passive system [201], the EP-sensitivity enhancement in PT systems is significant in the sense that due to the gain-loss balance, the supermodes in principle have much narrower linewidths near EP points, which thus yield a narrower spectral background. This narrow spectral background is in favor of measurements in practice. By exploiting non-Hermiticity to enhance sensing capabilities, the PT-microcavity systems pave the way towards a new class of chip-scale ultrasensitive sensors beyond the conventional ones.

5. Conclusion and outlook

As we indicated in the Introduction, this compact review is by no means a complete presentation of the many new results on PT-symmetric microcavity structures that have appeared and continues to appear in the literature. For example, in our presentation we did not discuss cavity optomechanical systems, where phonon lasing [82-84] and chaos [85] can be dramatically modified from conventional circumstances. Similarly, we did not review the effects of quantum noise in a PT system [124-127] where its presence could lead to significantly different physics as compared to that expected from semiclassical approaches. Finally, following the extension of PT symmetry to optics and photonics, there are some recent interesting studies in investigating whether topological interface states can exist at all in PT-symmetric photonic systems [66,100,130,202-204]. After two decades effort, it seems to us at this moment that PT symmetry in the classical settings is fairly well understood in theory. The extension of the subject to other research fields, such as AMO physics, condensed matter physics, and quantum optics will not only enrich the scope of theoretical analysis but also open doors for a variety of potential applications. Despite the rapid progress made in recent years, it is believed that many significant works are yet to appear. The interdisciplinary research will lead to discoveries within the broad class of PT symmetry. From this point of view, our paper is just a starting point for future research on the phenomenon. We will not be surprised as new observations on PT symmetry to appear.

Acknowledgments

We are indebted to our colleagues Xiao Y, Huang Y-P, Huang S-W, Xia F, He B, Li G-X an d Stokes K for fruitful collaboration and for enlightening discussions on several topics of the current review. The work of J W was supported by the NSF EFMA-1741693, NSF 1806519, and KSU. X J and M X were supported by the National Basic Research Program of China (2012CB921804), the National Key Research and Development Program (2016YFA0302500), the National Natural Science Foundation of China (Nos. 61435007, 11574144 and 11321063), the Fundamental Research Funds for the Central Universities. L J acknowledges the funding support from the ARO MURI, ARL, AFOSR MURI, Alfred P Sloan Foundation and the David and Lucile Packard Foundation.

References

[1] Vahala K J 2003 *Nature* **424** 839-46
[2] Vahala K 2004 *Optical Microcavities* (Singapore: World Scientific Publishing)
[3] Heebner J, Grover R and Ibrahim T 2008 *Optical Microresonators: Theory, Fabrication, and Applications* (London: Springer-Verlag)
[4] Chang L, Jiang X, Hua S, Chao Y, Wen J, Jiang L, Li G, Wang G and Xiao M 2013 *Nature Photon.* **8** 524-9
[5] Peng B, Özdemir S K, Lei F, Monifi F, Gianfreda M, Long G L, Fan S, Nori F, Bender C M and Yang L 2014 *Nat. Phys.* **10** 394-8
[6] Hodaei,H, Miri M A, Heinrich M, Christodoulides D N and Khajavikhan M 2014 *Science* **346** 975-8
[7] Feng L, Wong Z J, Ma R, Wang Y and Zhang X 2014 *Science* **346** 972-5
[8] Wong Z J, Xu Y-L, Kim J, O'Brien K, Wang Y, Feng L and Zhang X 2016 *Nature Photon.* **10** 796-801
[9] Hodaei H, Hassan A U, Wittek S, Garcia-Gracia H, El-Ganainy R, Christodoulides D N and Khajavikhan M 2017 *Nature* **548** 187-91
[10] Heiss W D 2012 *J. Phys. A* **45** 444016
[11] Rotter I 2009 *J. Phys. A* **42** 153001
[12] Rotter I and Bird J P 2015 *Rep. Prog. Phys.* **78** 114001
[13] Heiss W D 2004 *J. Phys. A* **37**, 2455-64
[14] Berry M V 2004 *Czech. J. Phys.* **54** 1039-47
[15] Dembowski C, Dietz B, Gräf H-D, Harney H L, Heine A, Heiss W D and Richter A 2004 *Phys. Rev. E* **69** 056216
[16] Kato T 1980 *Perturbation Theory for Linear Operators* (New York: Springer-Verlag)
[17] Bender C M, Boettcher S and Meisinger P N 1999 *J. Math. Phys.* **40** 2201-29
[18] Bender C M 2005 *Contemporary Phys.* **46** 277-92
[19] Bender C M 2007 *Rep. Prog. Phys.* **70** 947-1018

[20] Makris K G, El-Ganainy R, Christodoulides D N and Musslimani Z H 2011 *Int. J. Theor. Phys.* **50** 1019-41
[21] Zyablovsky A A, Vinogradov A P, Pukkov A A, Dorofeenko A V and Lisyansky A A 2014 *Physics – Uspekhi* **57** 1063-82
[22] Kottos T and Aceves A B 2016 *Contemporary Optoelectronics* Ed. Shulika O and Sukhoivanov I (Dordrecht: Springer)
[23] Konotop V V, Yang J and Zezyulin D A 2016 *Rev. Mod. Phys.* **88** 035002
[24] Suchkov S V, Sukhorukov A A, Huang J, Dmitriev S V, Lee C and Kivshar Y S 2016 *Laser Photon. Rev.* **10** 177-213
[25] Feng L, El-Ganainy R and Ge L 2017 *Nature Photon.* **11** 752-62
[26] El-Ganainy, Makris K G, Khajavikhan M, Musslimani Z H, Rotter S and Christodoulides D N 2018 *Nat. Phys.* **14** 11-19
[27] Zhang Z, Ma D, Sheng J, Zhang Y, Zhang Y-P and Xiao M 2018 *J. Phys. B: At. Mol. Opt. Phys.* **51** 072001
[28] Schindler J, Lin Z, Lee J M, Ramezani H, Ellis F M and Kottos T 2012 *J. Phys. A* **45** 444029
[29] Mostafazadeh A 2010 *Int. J. Geom. Methods Mod. Phys.* **7** 1191-306
[30] Geyer H, Heiss D and Znojil M 2006 Eds. *J. Phys. A* Special issue Vol. **39** No. 32
[31] Fring A, Jones H F and Znojil M 2008 Eds. *J. Phys. A* Special issue Vol. **41** No. 24
[32] Bender C M, Fring A, Günther U and Jones H F 2012 Eds. *J. Phys. A* Special issue Vol. **44**
[33] Bender C M and Boettcher S 1998 *Phys. Rev. Lett.* **80** 5243-6
[34] Lee T D and Wick G C 1969 *Nucl. Phys. B* **9** 209-43
[35] Ruschhaupt A, Delgado F and Muga J G 2005 *J. Phys. A* **38** L171-6
[36] El-Ganainy R, Makris K G, Christodoulides D N and Musslimani Z H 2007 *Opt. Lett.* **32** 2632-4
[37] Makris K G, El-Ganainy R, Christodoulides D N and Musslimani Z H 2008 *Phys. Rev. Lett.* **100** 103904
[38] Musslimani Z H, Makris K G, El-Ganainy R and Christodoulides D N 2008 *Phys. Rev. Lett.* **100** 030402
[39] Guo A, Salamo G J, Duchesne D, Morandotti R, Volatier-Ravat M, Aimez V, Siviloglou G A and Christodoulides D N 2009 *Phys. Rev. Lett.* **103** 093902
[40] Rüter C E, Makris K G, El-Ganainy R, Christodoulides D N, Segev M and Kip D 2010 *Nat. Phys.* **6** 192-5
[41] Schindler J, Li A, Zheng M C, Ellis F M and Kottos T 2011 *Phys. Rev. A* **84** 040101(R)
[42] Bender N, Factor S, Bodyfelt J D, Ramezani H, Christodoulides D N, Ellis F M and Kottos T 2013 *Phys. Rev. Lett.* **110** 234101
[43] Assawaworrarit S, Yu X and Fan S 2017 *Nature* **546** 387-90
[44] Bittner S, Dietz B, Günther U, Harney H L, Miski-Oglu M, Richter A and Schäfer F 2012 *Phys. Rev. Lett.* **108** 024101
[45] Bender C M, Berntson B K, Parker D and Samuel E 2013 *Am. J. Phys.* **81** 173-9
[46] Chtchelkatchev N M, Golubov A A, Baturina T I and Vinokur V M 2012 *Phys. Rev. Lett.* **109** 150405
[47] Regensburger A, Bersch C, Miri M-A, Onishchukov G, Christodoulides D N and Peschel U 2012 *Nature* **488** 167-71
[48] Regensburger A, Miri M-A, Bersch C, Näger J, Onishchukov G, Christodoulides D N and Peschel U 2013 *Phys. Rev. Lett.* **110** 223902
[49] Wimmer M, Regensburger A, Miri M-A, Bersch C, Christodoulides D N and Peschel U 2015 *Nat. Commun.* **6** 7782
[50] Feng L, Xu Y-L, Fegadolli W S, Lu M-H, Oliveira J E B, Almeida V R, Chen Y-F and Scherer A 2013 *Nat. Mater.* **12** 108-13

[51] Lawrence M, Xu N, Zhang X, Cong L, Han J, Zhang W and Zhang W 2014 *Phys. Rev. Lett.* **113** 093901
[52] Sun Y, Tan W, Li H-Q, Li J and Chen H 2014 *Phys. Rev. Lett.* **112** 143903
[53] Gu Z, Zhang N, Lyu Q, Li M, Xiao S and Song Q 2016 *Laser Photon. Rev.* **10** 588-94
[54] Wimmer M, Miri M-A, Christodoulides D N and Peschel U 2015 *Sci. Rep.* **5** 17760
[55] Xu Y-L, Fegadolli W S, Gan L, Lu M-H, Liu X-P, Li Z-Y, Scherer A and Chen Y-F 2016 *Nat. Commun.* **7** 11319
[56] Yan Y and Giebink N C 2014 *Adv. Opt. Mater.* **2** 423-7
[57] Jia Y, Yan Y, Kesava S V, Gomez E D and Giebink N C 2015 *ACS Photon.* **2** 319-25
[58] Hahn C, Choi Y, Yoon J W, Song S H, Oh C H and Berini P 2016 *Nat. Commun.* **7** 12201
[59] Heinrich M, Miri M-A, Stützer S, El-Ganainy R, Nolte S, Szameit A and Christodoulides D N 2014 *Nat. Commun.* **5** 3698
[60] Fleury R, Sounas D and Alù A 2015 *Nat. Commun.* **6** 5905
[61] Shi C, Dubois M, Chen Y, Cheng L, Ramezani H, Wang Y and Zhang X 2016 *Nat. Commun.* **7** 11110
[62] Aurégan Y and Pagneux V 2017 *Phys. Rev. Lett.* **118** 174301
[63] Zhang Z, Zhang Y, Sheng J, Yang L, Miri M-A, Christodoulides D N, He B, Zhang Y and Xiao M 2016 *Phys. Rev. Lett.* **117** 123601
[64] Peng P, Cao W, Shen C, Qu W, Wen J, Jiang L and Xiao Y 2016 *Nat. Phys.* **12** 1139-45
[65] Hodaei H, Miri M-A, Hassan A U, Hayenga W E, Heinrich M, Christodoulides D N and Khajavikhan M 2016 *Laser Photon. Rev.* **10** 494-9
[66] Weimann S, Kremer M, Plotnik Y, Lumer Y, Nolte S, Makris K G, Segev M, Rechtsman M C and Szameit A 2017 *Nat. Mater.* **16** 433-8
[67] Benisty H et al 2011 *Opt. Express* **19** 18004-19
[68] Lupu A T, Benisty H and Degiron A 2013 *Opt. Express* **21** 21651-68
[69] Alaeian H and Dionne J A 2014 *Phys. Rev. B* **89** 075136
[70] Huang C M, Ye F W, Kartashov Y V, Malomed B A and Chen X F 2014 *Opt. Lett.* **39** 5443-6
[71] Lazarides N and Tsironis G 2013 *Phys. Rev. Lett.* **110** 053901
[72] Castaldi G, Savoia S, Galdi V, Alú A and Engheta N 2013 *Phys. Rev. Lett.* **110** 173901
[73] Fleury R, Sounas D L and Alù A 2014 *Phys. Rev. Lett.* **113** 023903
[74] Lien J, Chen Y, Ishida N, Chen H, Hwang C and Nori F 2015 *Phys. Rev. B* **91** 024511
[75] Chestnov I Yu, Demirchyan S S, Alodjants A P, Rubo Y G and Kavokin A V 2016 *Sci. Rep.* **6** 19551
[76] Klaiman S, Günther U and Moiseyev N 2008 *Phys. Rev. Lett.* **101** 080402
[77] Cartarius H and Wunner G 2012 *Phys. Rev. A* **86** 013612
[78] Kartashov Y V, Konotop V V and Zezyulin D A 2014 *Europhys. Lett.* **107** 50002
[79] Rubinstein J, Sternberg P and Ma Q 2007 *Phys. Rev. Lett.* **99** 167003
[80] Lee J M, Kottos T and Shapiro B 2015 *Phys. Rev. B* **91** 094416
[81] Nazari F, Bender N, Ramezani H, Moravvej-Farshi M K, Christodoulides D N and Kottos T 2014 *Opt. Express* **22** 9574-84
[82] Jing H, Ozdemir S K, Lü X-Y, Zhang J, Yang L and Nori F 2014 *Phys. Rev. Lett.* **113** 053604
[83] Xu X-W, Liu Y, Sun C-P and Li Y 2015 *Phys. Rev. A* **92** 013852
[84] He B, Yang L and Xiao M 2016 *Phys. Rev. A* **94** 031802(R)
[85] Lü X-Y, Jing H, Ma J-Y and Wu Y 2015 *Phys. Rev. Lett.* **114** 253601
[86] Kepesidis K V, Milburn T J, Huber J, Makris K G, Rotter S and Rabl P 2016 *New J. Phys.* **18** 095003
[87] Zhu X, Ramezani H, Shi C, Zhu J and Zhang X 2014 *Phys. Rev. X* **4** 031042
[88] Nazari F, Nazari M and Morawej-Farshi M K 2011 *Opt. Lett.* **36** 4368-70
[89] Li K and Kevrekidis P G 2011 *Phys. Rev. E* **83** 066608
[90] Joglekar Y N, Scott D, Babbey M and Saxena A 2010 *Phys. Rev. A* **82** 030103(R)

[91] Bendix O, Fleischmann R, Kottos T and Shapiro B 2009 *Phys. Rev. Lett.* **103** 030402
[92] He B, Yang L, Zhang Z and Xiao M 2015 *Phys. Rev. A* **91** 033830
[93] Suchkov S V, Dmitriev S V, Malomed B A and Kivshar Y S 2012 *Phys. Rev. A* **85** 033825
[94] Suchkov S V, Malomed B A, Dmitriev S V and Kivshar Y S 2011 *Phys. Rev. A* **84** 046609
[95] Ramezani H, Kottos T, El-Ganainy R and Christodoulides D N 2010 *Phys. Rev. A* **82** 043803
[96] Longhi S 2009 *Phys. Rev. Lett.* **103** 123601
[97] Longhi S 2009 *Phys. Rev. B* **80** 165125
[98] Longhi S 2010 *Phys. Rev. A* **81** 022102
[99] Mostafazadeh A and Mehri-Dehnavi 2009 *J. Phys. A* **42** 125303
[100] Schomerus H 2013 *Opt. Lett.* **38** 1912-4
[101] Longhi S 2013 *Phys. Rev. A* **88** 052102
[102] Horsley S A R, Artoni M and La Rocca G C 2015 *Nat. Photon.* **9** 436-9
[103] Wasak T, Szańkowski P and Konotop V V 2015 *Opt. Lett.* **40** 5291-4
[104] El-Ganainy R, Dadap J I and Osgood R M Jr 2015 *Opt. Lett.* **40** 5086-9
[105] Zhong Q, Ahmed A, Dadap J I, Osgood R M Jr and El-Ganainy R 2016 *New J. Phys.* **18** 125006
[106] Malzard S, Poli C and Schomerus H 2015 *Phys. Rev. Lett.* **115** 200402
[107] Hang C, Huang G and Konotop V V 2013 *Phys. Rev. Lett.* **110** 083604
[108] Sheng J, Miri M-A, Christodoulides D N and Xiao M 2013 *Phys. Rev. A* **88** 041803(R)
[109] Cerjan A, Raman A and Fan S 2016 *Phys. Rev. Lett.* **116** 203902
[110] Uzdin R, Mailybaev A and Moiseyev N 2011 *J. Phys. A* **44** 435302
[111] Miri M-A, Regensburger A, Peschel U and Christodoulides D N 2012 *Phys. Rev. A* **86** 023807
[112] Ramezni H, Kottos T, Kovanis V and Christodoulides D N 2012 *Phys. Rev. A* **85** 013818
[113] Lin Z, Ramezani H, Eichelkraut T, Kottos T, Cao H and Christodoulides 2011 *Phys. Rev. Lett.* **106** 213901
[114] Longhi S 2011 *J. Phys. A* **44** 485302
[115] Jones H F 2012 *J. Phys. A* **45** 135306
[116] Mostafazadeh A 2013 *Phys. Rev. A* **87** 012103
[117] Kulishov M, Jones H F and Kress B 2015 *Opt. Express* **23** 18694-711
[118] Longhi S 2010 *Phys. Rev. A* **82** 031801(R)
[119] Chong Y D, Ge L and Stone A D 2011 *Phys. Rev. Lett.* **106** 093902
[120] Mostafazadeh A 2009 *Phys. Rev. Lett.* **102** 220402
[121] Longhi S 2010 *Phys. Rev. Lett.* **105** 013903
[122] Ge L, Chong Y D and Stone A D 2012 *Phys. Rev. A* **85** 023802
[123] Ambichl P, Makris K G, Ge L, Chong Y D, Stone A D and Rotter S 2013 *Phys. Rev. X* **3** 041030
[124] Schomerus H 2010 *Phys. Rev. Lett.* **104** 233601
[125] Yoo G, Sim H-S and Schomerus H 2011 *Phys. Rev. A* **84** 063833
[126] He B, Yan S-B, Wang J and Xiao M 2015 *Phys. Rev. A* **91** 053832
[127] Zhang L, Agarwal G S, Schleich W P and Scully M O 2017 *Phys. Rev. A* **96** 013827
[128] Szameit A, Rechtsman M C, Bahat-Treidel O and Segev M 2011 *Phys. Rev. A* **84** 021806
[129] Jović D M, Denz C and Belić M R 2012 *Opt. Lett.* **37** 4455-7
[130] Hu Y C and Hughes T L 2011 *Phys. Rev. B* **84** 153101
[131] Muga J G, Palao J P, Navarro B and Egusquiza I L 2004 *Phys. Rep.* **395** 357-428
[132] Cannata F, Dedonder J-P and Ventura A 2007 *Ann. Phys.* (Amsterdam) **322** 397-433
[133] Daley A J 2014 *Adv. Phys.* **63** 77-149
[134] Garcia S R, Prodan E and Putinar M 2014 *J. Phys. A* **47** 353001
[135] Cao H and Wiersig J 2015 *Rev. Mod. Phys.* **87** 61-111
[136] Moiseyev N 2011 *Non-Hermitian Quantum Mechanics* (Cambridge: Cambridge University Press)

[137] Gamow G 1928 *Z. Fur. Phys.* **51** 204-12
[138] Feshbach H, Porter C E and Weisskopf V F 1954 *Phys. Rev.* **96** 448-64
[139] Lindblad G 1976 *Commun. Math. Phys.* **48** 119-30
[140] Gorini V, Kossakowski A and Sudarshan E C G 1976 *J. Math. Phys.* **17** 821-5
[141] Feshbach H 1958 *Ann. Phys.* **5** 357-90
[142] Carmichael H 1993 *An Open Systems Approach to Quantum Optics* (Berlin: Springer)
[143] Rivas Á and Huelga S F 2012 *Open Quantum Systems: An Introduction* (Berlin: Springer)
[144] Barton G 1963 *Introduction to Advanced Field Theory* (Interscience Publishers)
[145] Wu T T 1959 *Phys. Rev.* **115** 1390-404
[146] Brower R C, Furman M A and Moshe M 1978 *Phys. Lett. B* **76** 213-9
[147] Harms B, Jones S and Tan C-I 1980 *Nucl. Phys. B* **171** 392-412
[148] Fisher M E 1978 *Phys. Rev. Lett.* **40** 1610-3
[149] Cardy J L 1985 *Phys. Rev. Lett.* **54** 1345-8
[150] Zamolodchikov A B 1991 *Nucl. Phys. B* **348** 619-41
[151] Caliceti E, Graffi S and Maioli M 1980 *Commun. Math. Phys.* **75** 51-66
[152] Haydock R and Kelly M J 1975 *J. Phys. C: Solid State Physics* **8** L290-3
[153] Hatano N and Nelson D R 1997 *Phys. Rev. B* **56** 8651-73
[154] Wigner E 1959 *Group Theory and Its Application to Quantum Mechanics of Atomic Spectra* (New York: Academic)
[155] Bender C M, Berry M V and Mandilara A 2002 *J. Phys. A* **35** L467-71
[156] Bender C M, Brody D C and Jones H F 2002 *Phys. Rev. Lett.* **89** 270401
[157] Bender C M, DeKieiet M and Klevansky 2013 *Philos. Trans. Royal Soc. A* **371** 20120523
[158] Bender C M 2016 *J. Phys. A* **49** 401002
[159] Mostafazadeh A 2002 *J. Math. Phys.* **43** 205-14
[160] Mostafazadeh A 2002 *J. Math. Phys.* **43** 2814-6
[161] Mostafazadeh A 2002 *J. Math. Phys.* **43** 3944-51
[162] Mostafazadeh A 2003 *J. Phys. A* **36** 7081-92
[163] Mostafazadeh A 2003 *J. Math. Phys.* **44** 974-89
[164] Mostafazadeh A 2004 *Czech. J. Phys.* **54** 1125-32
[165] Mostafazadeh A 2015 *Trends in Mathematics 145* (Switzerland: Springer International)
[166] Bender C M, Brody D C and Jones H F 2003 *Am. J. Phys.* **71** 1095-1102
[167] Brody D C *J. Phys. A* **47** 035305
[168] Schrödinger E 1926 *Ann. Phys.* (Leipzig) **79** 489-527
[169] Dragoman D and Dragoman M 1999 *Prog. Quantum Electron.* **23** 131-88
[170] Páre C, Gagnon L and Bélanger P-A 1992 *Phys. Rev A* **46** 4150-60
[171] Nienhuis G and Allen L 1993 *Phys. Rev A* **48** 656-65
[172] Krivoshlykov S G ed. 1994 *Quantum-Theoretical Formalism for Inhomogeneous Graded-Index Waveguides* (Berlin: Akademie Verlag)
[173] Man'ko V L 1986 in: *Lie Methods in Optics* eds Mondragón J S and Wolf K B (Berlin: Springer)
[174] Dragoman D and Dragoman M 2004 *Quantum-Classical Analogies* (Berlin: Springer)
[175] Yariv A and Yeh P 2007 *Photonics: Optical Electronics in Modern Communications* (Oxford: Oxford University Press)
[176] Haus H A 1984 *Waves and Fields in Optoelectronics* (New Jersey: Prentice-Hall)
[177] Dzyaloshinskii I E 1991 *Phys. Lett. A* **155** 62-4
[178] Canright G S and Rojo A G 1992 *Phys. Rev. Lett.* **68** 1601-4
[179] Peng B, Özdemir S K, Rotter S, Yilmaz H, Liertzer M, Monifi F, Bender C M, Nori F and Yang L 2014 *Science* **346** 328-32

[180] Kippenberg T J, Spillane S M and Vahala K J 2002 *Opt. Lett.* **27** 1669-71
[181] Phang S, Vukovic A, Creagh S C, Benson T M, Sewell P D and Gabriele G 2015 *Opt. Express* **23** 11493-507
[182] Hassan A U, Hodaei H, Miri M-A, Khajavikhan M and Christodoulides D N 2015 *Phys. Rev. A* **92** 063807
[183] Ge L and Tureci H E 2013 *Phys. Rev. A* **88** 053810
[184] Wu J-H, Artoni M and La Rocca G C 2015 *Phys. Rev. A* **91** 033811
[185] Wen J and Huang Y-P (unpublished)
[186] Yang F, Liu Y-C and You L 2017 *Phys. Rev. A* **96** 053845
[187] Graham R and Haken H 1968 *Z. Phys.* **210** 276-91
[188] Potton R J 2004 *Rep. Prog. Phys.* **67** 717-54
[189] Wen J *et al* 2015 *Photonics* **2** 498-508
[190] Jiang X *et al* 2016 *Sci. Rep.* **6** 38972
[191] Shi Y, Yu Z and Fan S 2015 *Nat. Photon.* **9** 388-92
[192] Ma J, Wen J, Hu Y, Jiang X, Jiang L and Xiao M 2018 arXiv: 1806.03169
[193] Hua S, Wen J, Jiang X, Hua Q, Jiang L and Xiao M 2016 *Nat. Commun.* **7** 13657
[194] Miao P, Zhang Z, Sun J, Walasik W, Longhi S, Litchinitsner N M and Feng L 2016 *Science* **353** 464-7
[195] Brandstetter M, Liertzer M, Deutsch C, Klang P, Schöberl J, Türeci H E, Strasser G, Unterrainer K and Rotter S 2014 *Nat. Commun.* **5** 4034
[196] Peng B *et al* 2016 *Proc. Natl Acad. Sci. USA* **113** 6845-50
[197] Chong Y D, Ge L, Cao H and Stone A D 2010 *Phys. Rev. Lett.* **105** 053901
[198] Wan W, Chong Y D, Ge L, Noh H, Stone A D and Cao H 2011 *Science* **331** 889-92
[199] Wiersig J 2014 *Phys. Rev. Lett.* **112** 203901
[200] Wiersig J 2016 *Phys. Rev. A* **93** 033809
[201] Chen W, Özdemir S K, Zhao G, Wiersig J and Yang L 2017 *Nature* **548** 192-6
[202] Yuce C 2015 *Phys. Lett. A* **379** 1213-8
[203] Harter A K, Lee T E and Joglekar Y N 2016 *Phys. Rev. A* **93** 062101
[204] Poli C, Bellec M, Kuhl U, Mortessagne F and Schomerus H 2015 *Nat. Commun.* **6** 6710